\documentclass[aps,prd,amsmath,nofootinbib,floatfix,fleqn]{revtex4}
\pdfoutput = 1

\usepackage{color}
\usepackage[dvipsnames]{xcolor}
\usepackage{graphicx}
\usepackage{bm}
\usepackage[utf8]{inputenc}
\usepackage{subfig}
 
 \usepackage{float}
 \usepackage{epsfig}% need for subequations
\usepackage{pdfpages}
 
\definecolor{darkblue}{rgb}{0.0,0,0.5}
\definecolor{darkgreen}{rgb}{0.0,0.3,0.0}
\definecolor{redish}{rgb}{0.675,0,0.2}
\definecolor{red}{rgb}{0.8,0,0}
\definecolor{green}{rgb}{0,0.6,0}
\definecolor{blue}{rgb}{0,0,0.8}

\usepackage[unicode=true, bookmarks=false, linkcolor = darkblue, citecolor = redish, breaklinks=false, colorlinks=true, hyperfootnotes=true]{hyperref}

\begin{document}

\date{\today}
\title{
Testing momentum dependence of the nonperturbative hadron structure\\ in a global QCD analysis}

\preprint{}

\author{Aurore Courtoy$^{(a,1)}$,  Pavel Nadolsky$^{(b,2)}$}

\affiliation{$^{(a)}$Instituto de F\'isica, Universidad Nacional Aut\'onoma de M\'exico\\
Apartado Postal 20-364, 01000 Ciudad de M\'exico, Mexico \\
$^{(b)}$Department of Physics, Southern Methodist University, Dallas, TX 75275-0181, U.S.A. }
\email{$^{1}$aurorecourtoy@gmail.com, $^{2}$nadolsky@smu.edu}

 \date{\today}

\begin{abstract}
     We discuss strategies for comparisons of nonperturbative QCD predictions for parton distribution functions (PDFs)  with high-energy experiments in the region of large partonic momentum fractions $x$. Analytic functional forms for PDFs cannot be uniquely determined solely based on discrete experimental measurements because of a mathematical property of mimicry of PDF parametrizations that we prove using a representation based on B\'ezier curves. Predictions of nonperturbative QCD approaches for the $x$ dependence of PDFs instead should be cast in a form that enables decisive comparisons against experimental measurements. Predictions for effective power laws of $(1-x)$ dependence of PDFs may play this role. Expectations for PDFs in a proton based on quark counting rules are compared against the effective power laws of $(1-x)$ dependence satisfied by CT18 next-to-next-to-leading order parton distributions. We comment on implications for studies of PDFs in a pion, in particular on the comparison of nonperturbative approaches with phenomenological PDFs.  
\end{abstract}

\maketitle

\vspace{.5cm}

\section{Introduction}
Quantum Chromodynamics (QCD) governs interactions of strongly interacting particles and predicts existence of bound states of hadronic matter. The scale dependence of the QCD coupling constant leads to the existence of two regimes of the strong interaction --the nonperturbative and perturbative -- that result in formation of hadronic bound states  at low energies and in quasi-free interactions of QCD partonic degrees of freedom at high energies. While both regimes require complex theoretical treatments, the perturbative regime of QCD has the advantage of rendering calculable predictions using the expansions in small parameters such as the inverse hard energy scale of the process and small QCD coupling constant. Experiments at the Large Hadron Collider and other facilities test the perturbative phase of QCD to high accuracy. 

As for the nonperturbative regime, powerful approaches characterize internal dynamics of hadrons based on the models of the hadronic wave function (or the bound-state amplitude) and the effective Lagrangian approaches incorporating emergent low-energy symmetries. The pioneering studies of the hadron structure in  nonperturbative models (see, {\it e.g.},~\cite{Jaffe:1974nj,Davidson:1994uv}) have paved the way for recent  rapid advancements, driven particularly by  discretized (lattice) QCD \cite{Lin:2017snn,Lin:2020rut} as well as by other approaches such as analytical representations ({\it e.g.}~\cite{Kumericki:2016ehc}) or Schwinger-Dyson formalism, {\it e.g.}~\cite{Roberts:1994dr}. Of particular interest to these studies are parton distribution functions (PDFs) -- universal functions quantifying probabilities for finding partons in a fast-moving hadron probed at a factorization scale $\mu \gg 1\mbox{ GeV}$.

The highly challenging computation of a PDF for an arbitrary parton becomes more amenable when the parton carries a significant fraction $x$ of the hadron's momentum, of order 0.1 or more. Recent nonperturbative/lattice computations provide many predictions for PDFs at $x\rightarrow 0.1$, as well as of the Mellin moments dominated by large-$x$ PDFs~\cite{Lin:2020rut}. At $x\gtrsim 0.5$, the flavor dependence of the proton wave function further simplifies, with only the up and down quarks having the appreciable PDFs. Furthermore, under specific conditions outlined in Sec.~\ref{sec:QCR}, notably requiring that the parton entering the hard scattering carries nearly all of the hadron's momentum (i.e., when $x\rightarrow 1$), the proton wave function right before the hard scattering may reduce to a small number of simplest quasi-free partonic Fock states, which in turn may allow one to predict the $x\to 1$ asymptotics of the PDFs probed at sufficiently high $\mu$. This physical picture gives rise to the famous quark counting rules~\cite{Brodsky:1973kr,Brodsky:1974vy,Lepage:1980fj}. The $x$ dependence of the PDFs at $x>0.5$, the topic of interest for this paper, may open avenues for appraising the predictiveness of  nonperturbative approaches. 

On the phenomenological side, the PDFs are used to predict long-distance contributions to hadronic cross sections, when combined with perturbative parton scattering cross sections. Precise  phenomenological parametrizations of PDFs \cite{Dulat:2015mca,Harland-Lang:2014zoa,Ball:2017nwa,Alekhin:2017kpj,Accardi:2016qay,Harland-Lang:2019pla,Bertone:2017bme,Manohar:2017eqh} for unpolarized protons are determined by performing the global QCD analysis of experimental measurements. A global QCD analysis is a large-scale study involving fits of parametrized PDFs to various experimental data sets in the framework of perturbative QCD (PQCD). Flexible functional forms for these PDFs are fitted to measured cross sections in diverse high-energy processes, such as deeply inelastic scattering and production of vector bosons and jets. 
We will focus primarily on unpolarized proton PDFs, as they 
are best constrained experimentally, although baryons are not the simplest particles from the nonperturbative point of view.

Phenomenological PDFs can provide (and have provided) useful guidance for models of nonperturbative dynamics, {\it e.g.} by identifying the energy scale where the model is applicable~\cite{Stratmann:1993aw,Traini:1997jz}.
However, comparisons of model and phenomenological PDFs must not be done uncritically. On the one hand, the most common $\overline{MS}$ PDFs enter a factorized {\it approximation} for the hadronic scattering cross section that is valid up to process-dependent power-suppressed terms. While the nonperturbative computations predict the PDFs in a free hadron, the initial-state hadrons in high-energy scattering processes are not truly free. They interact with other participating particles through soft QCD interactions. Scale-dependent PDFs provide a logarithmic approximation to the process of collinear parton showering in the initial state. The full radiation pattern also depends on particle masses and kinematic constraints. The QCD factorization formulae approximate the hadronic cross sections in simple inclusive processes in a way that accounts for soft and collinear contributions, and neglects numerically small mass terms. The relation of these formulae to the nonperturbative PDFs includes power-suppressed terms that are not controlled to the necessary extent. 

On the other hand, the PDFs enter the fitted cross sections through elaborate, flavor-dependent convolution integrals and are determined from complex experimental measurements. The global analysis of PDFs relies on the critical assumption of universality,  that the PDFs do not depend on the hard-scattering process. Multiple factors contribute to the final PDF uncertainty, as detailed, for instance, in the recent reviews~\cite{Gao:2017kkx,Kovarik:2019xvh}. The question then arises, to which extent the phenomenological PDF analyses can genuinely reproduce the features reflecting nonperturbative dynamics. 

To illustrate these issues, we will revisit a classical problem in the PDF analysis, determination of the power laws that govern the falloff of PDFs as $x$ approaches one. Quark counting rules (QCRs), reviewed in Sec.~\ref{sec:QCR}, are one of the earliest predictions that the valence PDFs in the proton fall off roughly like $(1-x)^3$ both for the up and down quarks. Remarkably, the power law predicted by the QCRs, as well as the asymptotic behavior of the $d(x)/u(x)$ at $x\rightarrow 1$ --- a consequence of the spin-flavor extension of the original QCRs~\cite{Farrar:1975yb} --- are consistent with the behavior of the actual phenomenological PDFs, although alternative behaviors are not ruled out.

We will examine this $(1-x)$ falloff  in the recent CT18 next-to-next-to-leading order global analysis \cite{Hou:2019efy}. (For an informative study of the empirical small-$x$ and large-$x$ power laws in the NNPDF fit, which follows a different methodology, see Ref.~\cite{Ball:2016spl}.)
We start in Sec.~\ref{sec:QCRweakconstant} by briefly reviewing the rationale for the QCRs and by stating the conditions under which the QCRs are expected to hold. Next, we revisit the connection between the PDFs and factorized formulas for fitted hadronic cross sections in Sec.~\ref{sec:LargeXRealQCD}. 

Can the power laws predicted by the QCRs be tested by experimental hadron measurements? We address this question by presenting a mathematical argument in Sec.~\ref{sec:Mimicry} to show that the polynomial functional form of the structure functions or PDFs cannot be uniquely determined from experimental observations. This conclusion follows from basic properties of polynomial interpolation. As an alternative to the reconstruction of the analytic form, we introduce an empirical power law exponent for the proton PDFs, as defined in Eq.~(\ref{eq:A2Def}). The empirical exponent allows a phenomenologist to reliably confront theoretical models with the observed behavior of hadronic cross sections. We address the dependence of the empirical power law exponent on the functional form of the PDFs, factorization scale, and the type of the scattering process in the remainder of Sec.~\ref{sec:TestingQCRs}.

Functional forms of phenomenological PDFs incorporate various assumptions about the asymptotic behaviors of the PDFs at $x\rightarrow 0$ and $1$, including flavor dependence. It is important to understand compatibility of these assumptions with the experimental data. This study also serves as a sandbox problem illustrating broad aspects of comparisons of phenomenological PDFs with the large-$x$ predictions. In particular, investigations of the $x$ dependence of the PDFs in pions and kaons have been proposed as a powerful test to understand mechanisms for the emergence of the hadronic mass~\cite{Aguilar:2019teb}. 
Yet modern pion PDF analyses~\cite{Barry:2018ort,Novikov:2020snp} arrive at varied  conclusions about the validity of QCRs for the pion, or they may even appear to be at odds with expectations from nonperturbative approaches \cite{Holt:2010vj}. In  Sec.~\ref{sec:disc} we comment on the reconciliation of the physical pictures arising from the perturbative and nonperturbative descriptions of QCD and on implications for pion PDF studies.  

\section{Quark counting rules and QCD factorization}
\label{sec:QCR}

\subsection{A weakly coupled gauge theory \label{sec:QCRweakconstant}}

\begin{figure}[tb]
\centering
\includegraphics[width=1\textwidth]{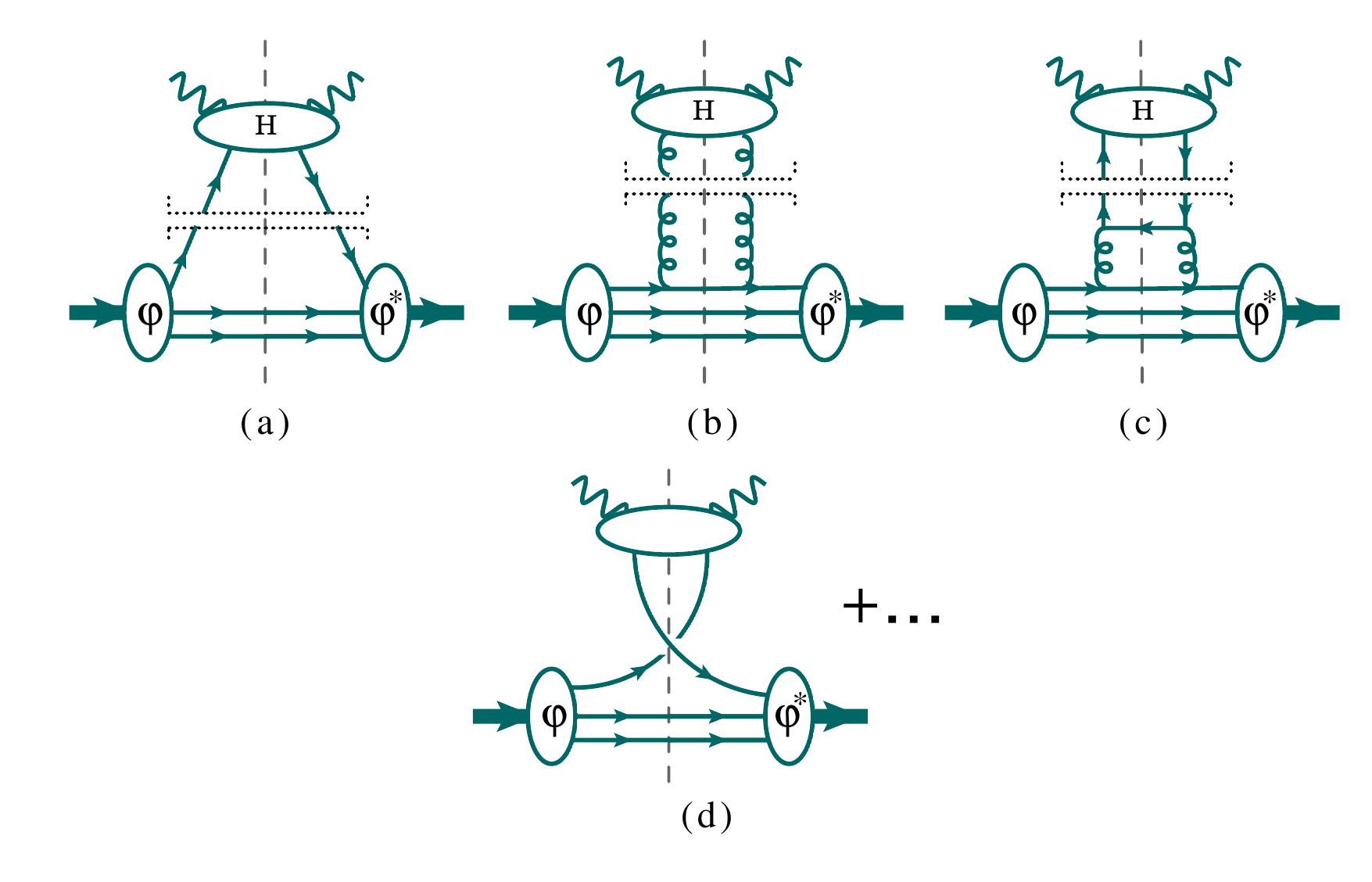}
\caption{Leading QCD radiative contributions 
giving rise to the counting rules for (a) valence quarks, (b) gluons, and (c) sea quarks. We assume $x\to 1$ and a very small QCD coupling constant. (d) A "resolved photon" diagram that is non-negligible for small virtualities of the photon.}
\label{fig:QCRdiagrams}
\end{figure}

\subsubsection{ Quark counting rules (QCRs) for structure functions} 
The QCRs for a structure function $F(x_B,Q^2)$ in  lepton-hadron deeply inelastic scattering arise from the parton model in gauge theories with small quark-boson coupling constants, such as QED or asymptotically free QCD. Consider a Feynman diagram in Fig.~\ref{fig:QCRdiagrams}(a) in such a weakly coupled theory with massless quarks. The diagram corresponds to scattering of a virtual photon $\gamma^*(q)$ on a highly boosted "proton" $p(P)$ whose lowest Fock state entering the hard scattering (at momentum resolution scales somewhat below $Q^2 \equiv -q^2$) consists of three weakly bound quarks. [Alternatively, we could consider scattering on a "meson" consisting of a quark and an antiquark.] $\phi$ is the low-energy (long-distance) part of the hadronic wave function, describing the binding of quarks into the hadron at virtualities much less than $Q^2$. $H$, the hard-scattering subgraph of the diagram, can be approximated by the quark-photon bag diagram (the squared tree-level amplitude of the $\gamma^*q$ scattering) if all couplings are small. 
The diagram in Fig.~\ref{fig:QCRdiagrams}(a) dominates the cross section when the $\gamma^*p$ scattering energy $W^2=Q^2\ \left(1/x_B -1\right)+m_p^2$ barely exceeds the mass $m_p^2$ of the initial proton. This regime corresponds to the maximal Bjorken variable, $x_{\rm B} \equiv  Q^2/(2P\cdot q) \to 1$. The contribution of this diagram to the structure function at $x_B \to 1$ behaves as
\begin{equation}
\label{eq:QCR_BF}
F_2(x_{\rm B},Q^2)\xrightarrow[x_{\rm B} \to 1] {} \left(1-x_{\rm B}\right)^{2n_s-1+2|\lambda_q-\lambda_A|} \cdot \left\{\mbox{const} + {\cal O}(1-x_{\rm B})\right\}
\;,
\end{equation}
with $n_s$ being the number of spectator partons (two for a baryon and one for a meson); and $\lambda_A$ and $\lambda_q$ denoting helicities of the parent hadron and active (struck) quark. For spin-averaged proton and pion structure functions, we obtain the limits 
\begin{equation}
\lim_{x_B\to 1}F_2^p(x_{\rm B},Q^2)\propto (1-x_{\rm B})^3, \quad 
\lim_{x_B\to 1}F_2^\pi(x_{\rm B},Q^2)\propto (1-x_{\rm B})^2 \label{eq:QCR_F2p}.
\end{equation}

The $(1-x_B)$ power law for $F_2(x_B,Q^2)$ thus arises when the $(n_s+1)$-quark Fock state dominates in the $x_B\to 1$ limit. In this picture, the $(1-x)$ falloff is driven primarily by semi-hard gluon propagators binding the $(n_s+1)$ quarks before the hard scattering,  on the top of long-distance binding effects included in the nonperturbative wave function $\phi$. The QCRs were initially demonstrated based on the examination of leading perturbative diagrams \cite{Ezawa:1974wm,Farrar:1975yb,Berger:1979du} as well as analyticity of partial-wave amplitudes \cite{Soper:1976jc}, and including helicity dependence as in Eq.~(\ref{eq:QCR_BF}) \cite{Farrar:1975yb}. They are also expected to apply in various nonperturbative approaches, see examples in  Sec.~\ref{sec:LargeXRealQCD}.  
Adding even more gluon propagators to the graph in Fig.~\ref{fig:QCRdiagrams}(a) suppresses the rate both by additional powers of $(1-x_B)$ and by additional factors of the coupling constant. The term of order $(1-x_B)$ in curly brackets in Eq.~(\ref{eq:QCR_BF}) arises from these higher-order radiative contributions. The extra suppression power can be found by counting the added propagators. Such contributions also introduce anomalous dimensions that make the $(1-x_B)$ exponent dependent on the renormalization scale $Q$ \cite{Soper:1976jc,Goldberger:1976vp}. 

We thus see that the QCRs are mostly directly formulated for the structure functions; QCRs for the PDFs are discussed below.  Sec.~\ref{sec:Mimicry} argues on general grounds that the power law introduced by the QCRs in Eq.~(\ref{eq:QCR_F2p}) cannot be directly tested. Instead, we construct an effective power-law exponent that can be compared against experimental measurements, as discussed in Sec.~\ref{sec:A2eff}. In the case of proton structure functions, Sec.~\ref{sec:a2effF2p} demonstrates that the effective power-law exponent predicted by the QCRs is compatible with the global QCD analysis of hadronic scattering data. See, in particular, the discussion of Fig.~\ref{fig:OverTheRainbow}.

\subsubsection{QCRs for electromagnetic form factors} 
\label{subsec:ff}
The described physics picture also applies to exclusive processes~\cite{Brodsky:1973kr,Brodsky:1974vy,Lepage:1980fj}.  Drell-Yan--West duality~\cite{Drell:1969km,West:1970av} relates deep inelastic structure functions near the threshold, $x_B\to 1$, to elastic electromagnetic form factors at large momentum transfer, $Q^2 \gg 1\mbox{ GeV}^2$. The scattering contributions that dominate the large-$x_B$ limit of the DIS structure function are thus expected to drive the $1/Q^2$ falloff of the elastic form factor~\cite{Soper:1976jc,Ezawa:1974wm}. For example, in the parton model, the inelastic structure function $\nu W_2(x_B)$ and electromagnetic form factor $F_1(Q^2)$, when both represented in terms of the target wave function, are related through position-dependent parton distributions which have an explicit parton density interpretation~\cite{Soper:1976jc}, implying the interdependent power laws for $\nu W_2(x_B)$ and  $F_1(Q^2)$. 
More generally  the structure functions and form factors are related  via impact-parameter generalized parton distributions for the $x$ region for which DGLAP evolution applies (see, {\it e.g.}, Ref.~\cite{Diehl:2003ny}).

\subsubsection{QCRs for parton distribution functions: DIS scheme} 
In real-life QCD, computation of high-energy hadronic cross sections involves factorization of long-distance and short-distance QCD radiative contributions. The power law fall-off of factorized structure functions translates into the fall-off of PDFs $f_a(x,Q^2)$ at large light-cone momentum fractions $x$ (which differ from the Bjorken variable $x_{\rm B}$ starting at the next-to-leading order). 

The connection is most transparent in {\bf the DIS factorization scheme} \cite{Diemoz:1987xu}, where the neutral-current DIS structure function is given by the charge-weighted sum of quark parton distributions to all orders in the QCD coupling constant $\alpha_s$:
\begin{equation}
    \left.F_2(x_B,Q^2)\right|_{\rm DIS\ scheme}= \mbox{const}\cdot \sum_{i=u,d,...} e_i^2 \left(f_i^{\rm DIS}(x_B,Q^2)+f_{\bar i}^{\rm DIS}(x_B,Q^2)\right) + {\cal O}(M/Q).
\label{eq:QCR_PDF}
\end{equation}
For valence-quark DIS PDFs in the proton, this implies the same fall-off power as in Eqs.~(\ref{eq:QCR_BF}) and (\ref{eq:QCR_F2p}),
\begin{equation}
\lim_{x\to 1}f_{i/p}^{\rm DIS}(x_,Q^2)= (1-x)^{A_{2i}} \cdot \left\{\mbox{const} + {\cal O}(1-x)\right\},
\label{eq:QCR_A2i}
\end{equation}
where 
\begin{equation}
A_{2i}=3 \mbox{\quad for } i=u\mbox{ and }d,
\label{eq:QCR_valence}
\end{equation}
if the diagram in Fig.~\ref{fig:QCRdiagrams}(a) dominates. Note that the predicted falloff power is the same for valence up and down quarks. 

Starting at the next order in $\alpha_s$, the cross sections receive significant contributions from QCD radiation. DGLAP evolution equations \cite{Gribov:1972ri,Lipatov:1974qm,Altarelli:1977zs,Dokshitzer:1977sg}  implement a collinear approximation for initial-state QCD radiation, valid when $Q^2$ is much larger than $1\mbox{ GeV}^2$. The gluon and sea (anti)quark PDFs are generated by collinear radiation off the valence-quark lines, with the respective lowest-order diagrams shown in Figs.~\ref{fig:QCRdiagrams}(b) and (c). The splitting functions for the $q\to g$ and $g\to q$ splittings in these diagrams are 
\begin{equation}
    P_{g\leftarrow q}(x)=\frac{\alpha_s}{2\pi}\ C_F\left(\frac{1 + (1-x)^2}{x}\right)+...\ , \quad\quad P_{q\leftarrow g}(x)=\frac{\alpha_s}{2\pi}\frac{1}{2}\left(x^2 + (1-x)^2\right)+...\ . \label{eq:Pgq}  
\end{equation}
By computing the convolutions of $P_{g\leftarrow q}(x)$ and $P_{q\leftarrow g}(x)$ with the leading term of the valence PDF, Eqs.~(\ref{eq:QCR_PDF}), \ref{eq:QCR_A2i}), one determines the lowest-order estimates for the falloff exponents for sea (anti)quark and gluon PDFs in the proton:
\begin{eqnarray}
A_{2i} &=& 4 \mbox{\quad for } i=g,\mbox{ Fig.~\ref{fig:QCRdiagrams}(b)}; \label{eq:QCR_gluon} \\ 
A_{2i} &=& 5 \mbox{\quad for } \bar i=\bar u, \bar d,...\ , \mbox{ Fig.~\ref{fig:QCRdiagrams}(c)}. \label{eq:QCR_sea}  
\end{eqnarray}

If, instead, we solve the DGLAP differential equations for the scale evolution, we exponentiate the cumulative effect of collinear splittings from all $\alpha_s$ orders. The solution introduces anomalous dimensions for the leading asymptotic powers $A_{2i}(Q^2)$. In QCD, the respective anomalous dimensions are positive \cite{Ball:2016spl}; $A_{2i}(Q^2)$ grow with $Q^2$.
For an arbitrary $Q^2$, the QCRs thus predict
\begin{equation}
\label{eq:QCR_A2iQ2}
A_{2i}(Q^2) \geq 3,\ 4,\ \mbox{ and } 5
\end{equation}
for the valence, gluon, and sea quark PDFs in the proton, respectively.

The errors in these estimates critically depend on the size of radiative corrections to the lowest Feynman diagrams in Figs.~\ref{fig:QCRdiagrams}(a-c). These come from the higher powers in the coupling constant as well as from the power-suppressed contributions of order $M/Q$,  as indicated in Eq.~(\ref{eq:QCR_PDF}), where $M$ is a nonperturbative scale of order 1 GeV. 

At $x\to 1$, the main channel of QCD radiation is due to emission of gluons off valence quarks, as described by the non-singlet DGLAP equation for the valence quark PDFs. Contributions with radiation off initial-state sea (anti)quarks and gluons are strongly suppressed by smallness of their respective PDFs. The $Q^2$ dependence of $A_{2i}(Q^2)$ computed based on the DGLAP equations thus reflects the magnitude of higher-order corrections beyond the simplest QCR picture. 

Figure~\ref{fig:QCRdiagrams}(d) presents an example of a contribution  that is normally not discussed in the derivations of the quark counting rules, but may be an important part of the inclusive $F(x_B, Q^2)$ at relatively low $Q^2$. In this diagram, the photon splits into a $q\bar q$ dipole that interacts with the hadronic state. For highly virtual photons, this contribution can be estimated perturbatively. At $Q^2$ of order $1\mbox{ GeV}^2$ or less, the resolved photon contribution is independent from Fig.~\ref{fig:QCRdiagrams}(a) and is not small. It requires an independent "resolved photon" PDF.

\subsubsection{QCRs for parton distribution functions: $\overline{MS}$ scheme} 
Modern phenomenological PDFs are provided predominantly in the $\overline{MS}$ factorization scheme, which offers a number of advantages compared to the DIS factorization scheme. The $\overline{MS}$ PDFs are defined in a process-independent way as summarized in the next subsection. In the $\overline{MS}$ scheme, the leading-power Feynman integrals for DIS inclusive cross sections are given by convolutions of perturbative Wilson coefficients $H_a$ and nonperturbative PDFs $f_{a/A}$ (where $A=p$ or $\pi$) according to Eq.~(\ref{eq:DISfactorization}). In Fig.~\ref{fig:QCRdiagrams}, factorization of Feynman subgraphs is indicated by the horizontal double dotted lines separating the hard and PDF parts. We expect roughly the same falloff powers for the $\overline{MS}$ PDFs as in Eq.~(\ref{eq:QCR_A2iQ2}). Differences between the DIS and $\overline{MS}$ factorization schemes start at the next-to-leading order in $\alpha_s$.

\subsection{Large-$x$ behavior of QCD processes \label{sec:LargeXRealQCD}}
We see that the QCRs reflect a simplified picture of hadron scattering, in which cross sections near the elastic limit are dominated by the lowest-order diagrams like those in Fig.~\ref{fig:QCRdiagrams}. Is this simple depiction tethered to realistic measurements? 

Our view is that an arbitrary hadron scattering process is likely to include substantial violations of the QCRs. Thus, the QCRs need not be precisely obeyed by all processes included in global PDF fits. However, there may be processes where the kinematics favors the dominance of the lowest diagrams, and the QCRs are more closely followed. Specifically, when the initial hadron is "minimally perturbed" by the hard scattering, the higher-order Fock states may be better suppressed in the elastic limit. Several factors may indicate the "minimally perturbed" regime, including the smallness of the QCD coupling constant and vanishing $Q^2$ dependence of the effective power laws preferred by the experimental measurement. 

The key assumption of the QCRs, that the $(1-x)$ dependence is determined mostly by scattering off a few quarks knocked out of the parent hadron, suggests two possible conditions under which the QCRs may hold.  
First, a weakly bound incoming state may be  required, so that only the diagram with the minimal number of semi-hard propagators gives an appreciable rate in the elastic limit. In that case, each additional perturbative vertex introduces a large suppression factor into the scattering rate. 

Second, the hadron-parton vertex  described by a bound-state amplitude $\phi$ reflects the low-energy dynamics, {\it i.e.} the long-distance interaction that cannot be approximated by a few (semi-)hard gluons. The respective part of the hadronic correlator function can be evaluated consistently in a fully nonperturbative approach to hadron binding, such as  MIT bag ({\it e.g.}~\cite{Jaffe:1974nj}), Isgur-Karl  ({\it e.g.}~\cite{Parisi:1976fz}) or  chiral quark soliton  models ({\it e.g.}~\cite{Diakonov:1996sr}) for the proton, Nambu--Jona-Lasinio or chiral quark models ({\it e.g.}~\cite{Davidson:1994uv})  as well as Schwinger-Dyson equations ({\it e.g.}~\cite{Hecht:2000xa}) for the pion. 
In such approaches, the proton correlator is computed starting with the lowest-energy bound  states consisting of three quark fields. The coupling here is large, but the analyticity of partial-wave amplitudes \cite{Soper:1976jc} indicates that the power-law falloff may be realized in a variety of theories that lead to asymptotic freedom at short distances.  The dominance of final Fock states with lowest parton multiplicities is essential for realizing the QCRs in both cases. In the latter case, the QCRs may be more evident in a subsample of DIS events in which the final-state hadron multiplicity is low. 

Exclusive processes like the deeply virtual Compton scattering  with a photon that minimally perturbs the proton, and with the QCD coupling constant and mass terms tuned down, should satisfy the QCRs, as shown in the original derivation of Brodsky and Farrar. In particular, we must assume that the QCD radiation is weak enough so that the excited intermediate Fock states with five or more partons (or equivalently, the correlator contributions with sea partons, or with disconnected topologies) are negligible. The QCRs do not directly hold if there are excited Fock states.

Neither picture -- a weak coupling or a minimally perturbed hadron-- applies automatically in typical experimental measurements used in the proton PDF fits. Indeed, a typical inclusive hadronic observable used to determine the phenomenological PDFs includes high-multiplicity events. But we may spot the trace of the QCRs in some kinematic regime, when final-state multiplicities are small, and the impact of other corrections is minimal. 

 We will further argue in Sec.~\ref{sec:pion} that the conditions supporting QCRs may be easier to achieve in pion scattering than in nucleon scattering. 

We take  neutral-current DIS on a proton as an example. In this process, two scales control the QCD radiation, the photon-proton center-of-mass energy $W^2$ (equal to the invariant mass squared of the hadronic final state) and the photon virtuality $Q^2$. For any reasonable choice of $W^2$ and $Q^2$ --- with the Bjorken regime limited by $W^2 > m_p^2$ with $W^2=m_p^2+(1-x_{\rm B})/x_{\rm B} \,Q^2$ --- the proton bound state is not minimally perturbed. There is no region of $W^2$ and $Q^2$ where both the QCD coupling $\alpha_s(Q^2)$ is small, and initial-state radiation into final states with more than three partons can be neglected.  Now consider three relevant kinematic regions of DIS:
\begin{itemize}
    \item In the elastic limit, {\it i.e.}, when $W^2 \rightarrow m_p^2 \sim 1 \mbox{ GeV}^2$, the proton mass $m_p^2$ is not negligible. The relevant three quark degrees of freedom are not massless and free, and some modification of the original Brodsky-Farrar motivation is necessary.
    \item  When $W^2$ increases up to about $4 \mbox{ GeV}^2$, the proton mass terms eventually become negligible, but the behavior of the DIS cross section is initially complex in this region because of the resonant contributions. Globally, one may expect that the picture based on scattering of quasi-free partons approximates the DIS cross section on average because of the Bloom-Gilman parton-hadron duality~\cite{Bloom:1970xb}. However, over small intervals of $W^2$, the cross section can exhibit very complex resonant behavior that does not satisfy the QCRs~\cite{Armstrong:2001xj,Liuti:2001qk}. 

    \item At even higher $W^2$, the power-suppressed terms become small. The {\it leading-power} contribution dominates a DIS structure function $F(x_{\rm B},Q^2)$ and can be factorized in terms of the PDFs $f_{a/p}$ and coefficient functions $H_{a}$ as
\begin{equation}
\label{eq:DISfactorization}
F(x_{\rm B}, Q^2)=\sum_a\int_{x_{\rm B}}^1 \frac{dx}{x}f_{a/p}(x,\mu^2)\, {H_{a}}\left(\frac{x_{\rm B}}{x},\frac{\mu^2}{Q^2}\right)
+ {\cal O}\!\left({M}/{Q}\right)
\;,
\end{equation}
where $H_{a}$ consists of a delta function for quark $a$ at the zeroth order of $\alpha_s$ and of respective higher-order radiative contributions for $a=q,g$ at higher orders.
The quark PDFs $f_{a/p}$, with their dependence on the partonic momentum fraction $x$ and factorization scale $\mu$, are defined in the $\overline{\rm MS}$ scheme as  
\begin{equation}
\label{eq:MSbarPDF}
f_{a/p}(x,\mu^2)=\frac{1}{4\pi} \int dy^- e^{-i x P^+y^-}
\left\langle P|\bar{\psi}_a(0,y^-, {\bf 0}) \gamma^+ W(y^-,0)\psi_a(0)|P\right\rangle
\;,
\end{equation}
where
\begin{equation}
\label{eq:WilsonLine}
W(y^-,0) = {\cal P} \exp\left(
-i g \int_0^{y^-}\!\!d\bar y^-  \widehat A^+(0^+,\bar y^-,\vec{0}_T)
\right)
\end{equation}
is the Wilson eikonal line, and we have used the light-cone coordinates, see {\it e.g.} Ref.~\cite{Kovarik:2019xvh}. 
At these $W^2$ and $Q^2$, we can finally talk about scattering on {\it nearly} independent initial-state partons, which nevertheless feel some long-distance interaction with other particles mediated by long-wavelength gluon fields. The factorization formula captures this interaction in two places, through the insertion of the eikonal line $W(y^-,0)$ in $f_{a/A}(x,\mu^2)$ to approximate the interaction of the initial-state quark field with the soft gluon field $\widehat A(y)$ connecting to the other particles, and through non-factorizable terms in the power-suppressed correction ${\cal O}(M/Q)$.

The inelastic cross section grows quickly in this region of $W^2$, indicating that final states with multiple partons are now easily produced. This effect is captured in the leading-power logarithmic approximation by the scale dependence of $f_{a/A}(x,\mu^2)$. These multi-parton final states violate the naive prediction of the QCRs. One indication of this violation is significant $Q^2$ dependence of the effective power law. 

\end{itemize}

{\bf Threshold resummation.} The collinear factorization formula (\ref{eq:DISfactorization}) is based on a highly non-trivial proof \cite{Collins:2011zzd,Collins:1998rz} that separates the leading-power convolution integral from power-suppressed terms ${\cal O}\!\left({M}/{Q}\right)$ such as target-mass corrections. The collinear formula is perturbatively stable when $W^2$ is of order $Q^2$. When $x\rightarrow 1$, the inclusive DIS cross section becomes sensitive to soft interactions among various particles that are not necessarily associated with the PDF(s). 
A different factorization formula, including a soft exponential factor, replaces the collinear factorization (\ref{eq:DISfactorization}) in this limit.  Soft radiation can be reliably approximated by a resummed all-order series of large logarithms if $Q$ is much larger than 1 GeV. At $Q$ of a few GeV, when the perturbative logarithms are not large, the threshold behavior is most sensitive to the nonperturbative part of the soft factor that should be fitted together with the PDFs.  In either case, radiation of {\it multiple} soft partons modifies the $x$ dependence of the DIS and DY cross sections at $x\to 1$ as compared to the QCR-based estimates. 

{\bf QCD factorization for other processes.}
To determine {\it phenomenological} functional forms for $\overline{MS}$ PDFs of various flavors, a global QCD analysis 
includes a comprehensive combination of experimental measurements in DIS, production of lepton pairs, jets, $t\bar t$ pairs, and other processes. As in the case of DIS, the connection between PDFs and inclusive cross sections relies on factorization theorems, and those are known with less confidence for more complex measurements. The cross sections used in the PDF fits are usually evaluated at a fixed order in $\alpha_s$ and often neglecting power-suppressed terms. 

The QCRs demonstrated for inclusive DIS cross sections do not translate automatically to the other processes. For example, while the leading-power collinear factorization for the Drell-Yan (DY) pair production cross section, 
\begin{equation}
\label{eq:DYfactorization}
\sigma ={} \sum_{a,b}\int\!dx_a \int\!dx_b\
f_{a/A}(x_a,\mu_\mathrm{F}^2)\, f_{b/B}(x_b,\mu_\mathrm{F}^2)
H_{a,b,x_a,x_b,\mu_\mathrm{F}^2}
+ {\cal O}\!\left({M}/{Q}\right)
\;,
\end{equation}
is structurally similar to that in inclusive DIS, as given in Eq.~(\ref{eq:DISfactorization}), the underlying scattering processes and factorization proofs are drastically different between two processes. 
In the Drell-Yan process, the underlying hadronic activity from multiperipheral scattering of {\it two} parent hadron remnants plays a far more prominent role and creates difficulties in proving the factorization~\cite{Collins:2011zzd}. The parent hadrons are more perturbed by soft interactions in hadron-hadron scattering than in DIS. Any factorization formula holds up to power-suppressed terms which are different in the collinear and threshold factorization formalisms, and which are different at some level in DIS and DY, or between different hadron and heavy nuclei parent species.\footnote{A well-known example of loss of universality of factorization are T-odd distributions in TMD factorization, which have opposite signs in DIS and DY process~\cite{Collins:2002kn}.} 

To summarize, in realistic QCD processes that determine the PDFs, the large-$x$ behavior is modified compared to the predictions of the massless parton model. The scope of modifications in the large-$x$ power laws introduced by higher Fock states, mass terms, resonant contributions, and nuclear effects varies by the scattering process. It is reasonable to expect large modifications of parton-model predictions in events with high final-state parton multiplicities.
Still, there can be situations that are close to realizing the assumptions that underlie the QCR's, such as when one tests the internal structure of a meson or looks at a subsample of DIS events with low final-state hadronic multiplicities.

\section{Testing large-$x$ PDFs in experimental measurements}
\label{sec:TestingQCRs}

    \subsection{B\'ezier curves as polynomial interpolations of discrete data\label{sec:Mimicry}}

Models of the hadron structure make concrete predictions for the $x$ dependence of the structure functions and PDFs. One can straightforwardly check the agreement of a given model with an experimental observation within the uncertainties. A stronger assertion, that the experiment {\it demands} the $1-x$ dependence of the PDFs to follow a specific power law, is difficult to demonstrate since the functional forms of the PDFs are not known exactly. This is clearly not possible in the presence of local or resonant structures that disagree with the global trend. Even when the PDF functional forms are restricted to be polynomial, the discrete experimental data can be compatible with multiple functional forms. 

To illustrate why, consider an idealized example, in which we seek a polynomial function $f^{(n)}(x)$ of degree $n$ to interpolate $k+1$ data points $\{x_0,p_0\},\{x_1,p_1\},$..., $\{x_k,p_k\}$ that have no uncertainty. Our points satisfy $0\leq x_i \leq 1$. From mathematics, we know that the existence and number of the interpolating solutions depend on the degree $n$ of the polynomial. 

If $n=k$, the unisolvence theorem guarantees that there exists a unique interpolating polynomial going through all points: $f^{(n)}(x_i)=p_i$. Two equivalent closed-form solutions for the interpolating polynomial are given by the Lagrange polynomial, 
\begin{equation}
{\cal L}^{(n)}(x) \equiv \sum_{i=0}^k p_i \prod_{\substack{m=1\\ m\neq i}}^k \frac{x-x_m}{x_i-x_m} \mbox{ for }n=k,
\label{eq:LagrangePolynomial}
\end{equation}
and by a B\'ezier curve of degree $n$,
\begin{equation}
{\cal B}^{(n)}(x) = \sum_{l=0}^n c_l\ B_{n,l} (x), 
\label{eq:BezierCurve}
\end{equation}
constructed from Bernstein basis polynomials
\begin{equation}
    B_{n,l} (x)\equiv \left(\begin{matrix} l  \\ n  \end{matrix}\right) x^l (1-x)^{n-l}.
\end{equation}

Denote the vector ${\cal B}^{(n)}(x_i)$ as $B$. This vector can be written in a matrix form  \cite{Farin:2001,Pomax:2020}, 
\begin{equation}
  B = T\cdot M \cdot C, 
\end{equation}
where $C\equiv\|c_l\|$; 
\begin{equation}
M\equiv \|m_{lp}\| \mbox{ with } m_{lp}=\begin{cases}
(-1)^{p-l}\left(
\begin{array}{c}
l\\
n
\end{array}\right)
\left(\begin{array}{c}
n-p\\
n-l
\end{array}\right), &  l\leq p\\
0, & l>p
\end{cases};
\label{M}
\end{equation}
and  $T\equiv \|t_{ip}\| $ with $t_{ip}=x_{i}^{p}$. Here $i$ runs from 0 to $k$, and $l,p$ run from 0 to $n$.

Given the matrix $P\equiv \|p_i\| $ of data values, the matrix $C$ for the B\'ezier curve ${\cal B}^{(n)}(x)$ going through all points satisfies \cite{Pomax:2020}
\begin{equation}
  C = M^{-1}\cdot T^{-1}\cdot P \mbox{ for } n=k. 
  \label{eq:CMTPsquare}
\end{equation}
This equation shows that $k+1$ data points uniquely determine the polynomial of order $n=k$, assuming no experimental errors. 

If $n<k$, an interpolating solution that goes through all points may not exist. Rather, there is a B\'ezier curve that minimizes the total squared distance to $p_i$, 
\begin{equation}
\chi^2(P,B) = \sum_{i=0}^k \left({\cal B}^{(n)}(x_i) - p_i\right)^2 = (P- T\cdot M\cdot C)^T \cdot (P- T\cdot M\cdot C).
\label{eq:chi2PB}
\end{equation}
The matrix of the coefficients of this B\'ezier curve is
\begin{equation}
  C = M^{-1}\cdot (T^T T)^{-1} \cdot T^T\cdot P \mbox{ for } n < k. 
  \label{eq:CMTPrectangular}
\end{equation}
The total squared distance from this special curve to $p_i$ is 
\begin{equation}
\min \chi^2(P,B) = P^T \cdot K^T \cdot K \cdot P \mbox{ for } n < k, 
\label{eq:chi2PBMin}
\end{equation}
where 
\begin{equation}
    K \equiv I_{(k+1)\times (k+1)} - T\cdot (T^T\cdot T)^{-1}\cdot T^T \quad.
    \label{eq:K}
\end{equation}

If we set $n=k$ in Eq.~(\ref{eq:CMTPrectangular}), it reduces to Eq.~(\ref{eq:CMTPsquare}). We also get $K=0$ and $\min \chi^2(P,B) =0$.

Finally, if $n>k$, an infinite number of polynomial solutions have $\min \chi^2(P,B) =0$. They can be constructed by adding $n-k$ arbitrary points to the Lagrange polynomial (\ref{eq:LagrangePolynomial}) found for $n=k$.

Equations (\ref{eq:CMTPsquare}) and (\ref{eq:CMTPrectangular}) for the coefficients of the B\'ezier curve are readily solvable and can be used to explore strategies for experimental determination of the $x$ dependence of the PDFs. The numerical solution for the interpolating polynomial is generally unstable for large $n$. The B\'ezier form is more stable compared to the other forms, sometimes allowing us to get stable interpolation for $n$ as large as 10 or 15. In our equations, numerical instabilities may arise from the inversion of matrix $T$ if $T$ is ill-conditioned when $n$ is large (especially if $k>n$), or when some points are spaced too closely in $x$. 

Once found from the data, the B\'ezier curve (\ref{eq:BezierCurve}) can be expanded into the monomial (Taylor) power series of $(1-x)$,
\begin{equation}
  {\cal B}^{(n)}(x) = \sum_{l=0}^n \bar{c}_l\  (1-x)^l,
  \label{eq:Bnonemx}
\end{equation}
which in turn can be compared against the predictions of the quark counting rules. The QCRs discussed in Sec.~\ref{sec:QCR} might predict that the empirically found coefficients $\bar{c}_l$ vanish when $l\leq A_2(Q)$. Or the whole set of $c_l$ or $\bar{c}_l$ can be compared against predictions of a given model. 

This comparison is impeded, however, by large cancellations between the terms with highest powers $l$ in the monomial expansion (\ref{eq:Bnonemx}) when $p_i$ values are sampled from a realistic PDF shape. The high-$l$ monomial terms tend to have alternating signs when interpolating such $p_i$. The monomial components with low $l$, signifying the cutoff by the QCRs and sensitive to the high-$l$ cancellations, vary significantly depending on the range and spacing of $x_i$.  

\begin{figure}[tb]
\centering
\includegraphics[width=.48\textwidth]{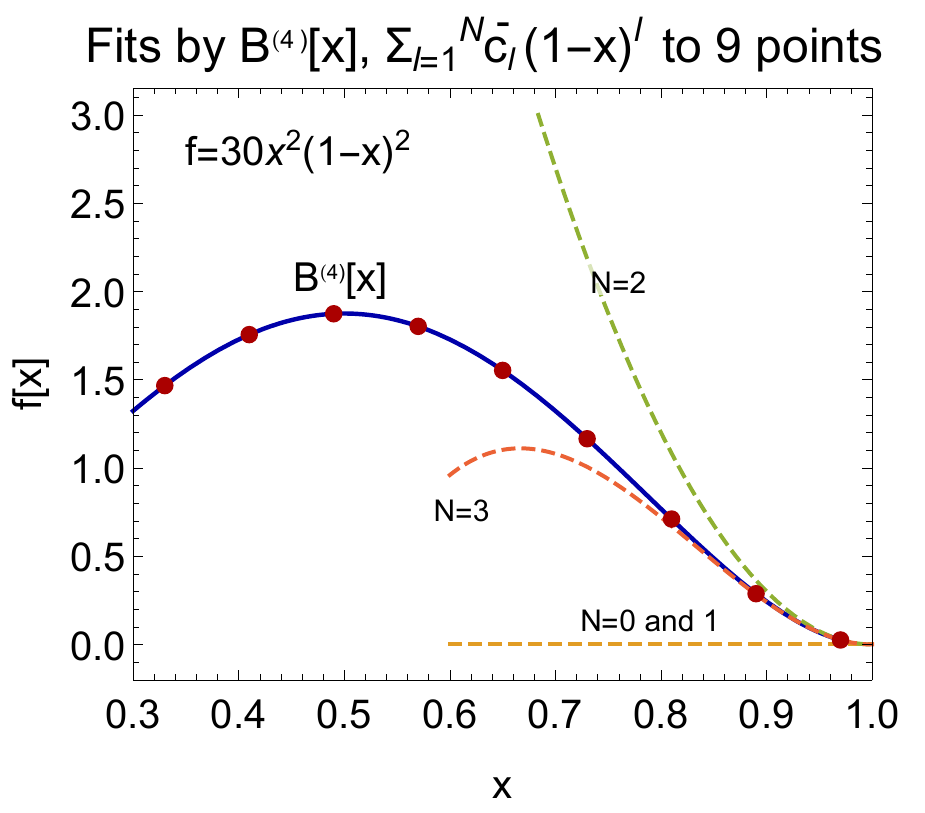}
\includegraphics[width=.48\textwidth]{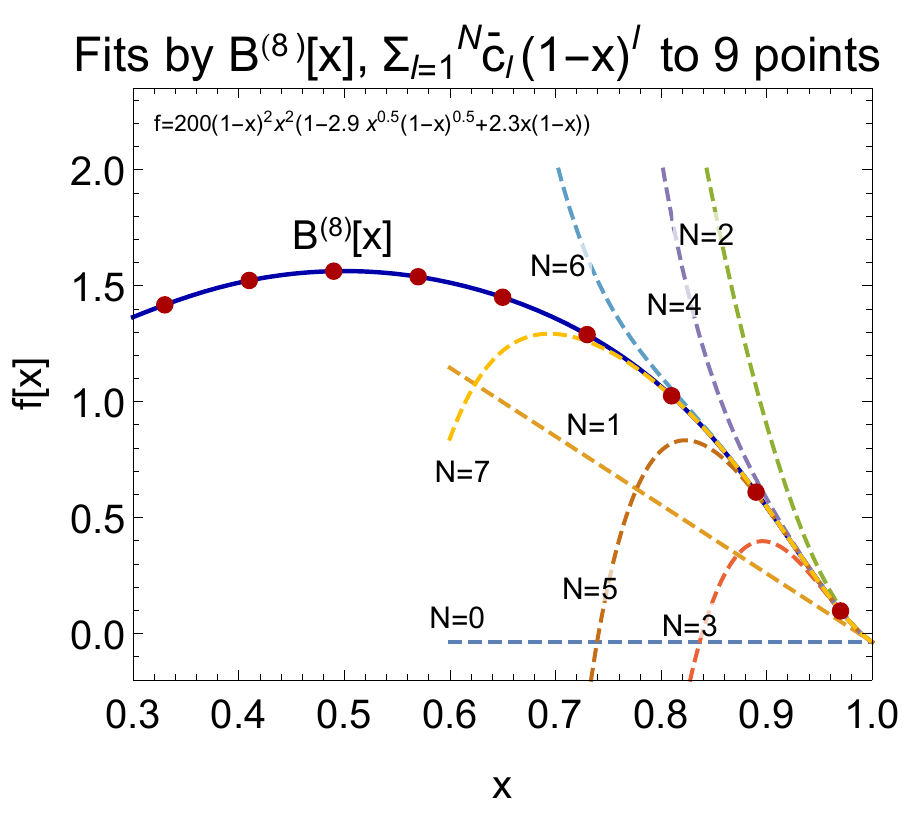}\\
\includegraphics[width=.48\textwidth]{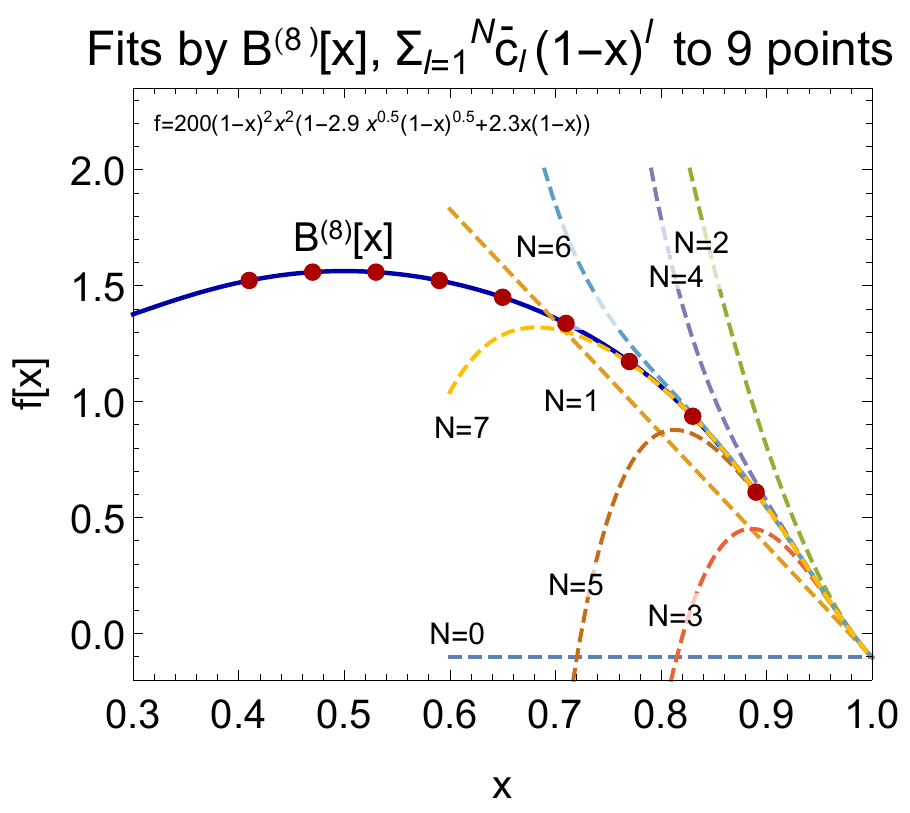}
\includegraphics[width=.48\textwidth]{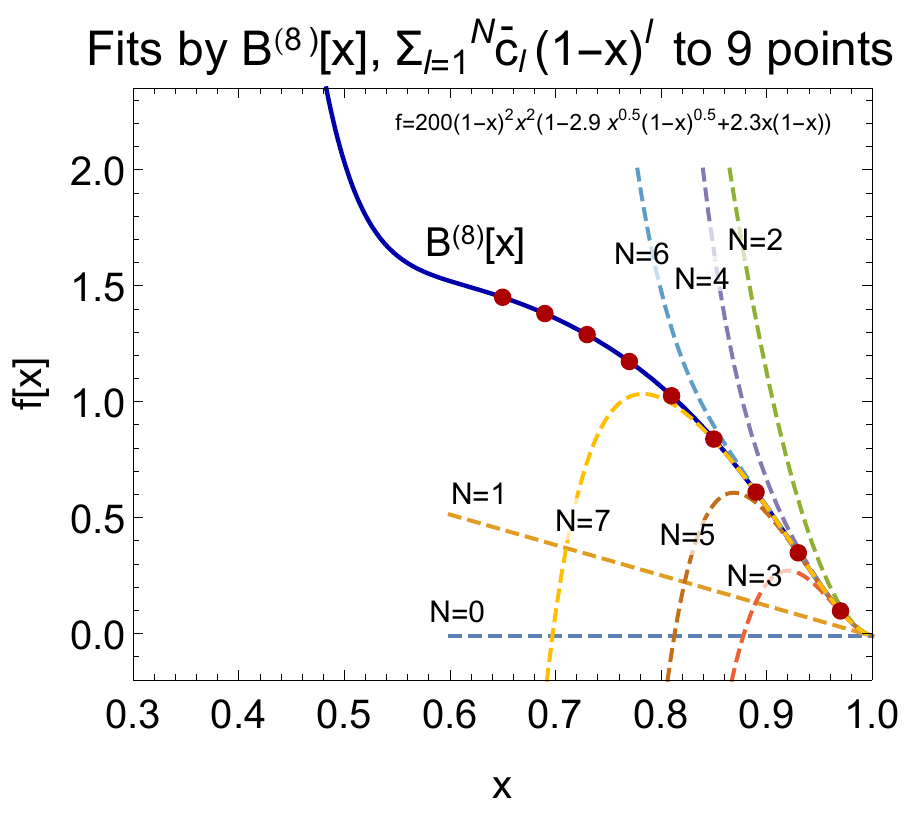}\\
\caption{(a,b) B\'ezier and polynomial fits to 9 discrete points sampled from the functions $f(x)$ specified in the figures. (c,d) Same as (b), for different ranges and spacings of $x$ covered by the sampled points.}
\label{fig:Beziers}
\end{figure}

Figure~\ref{fig:Beziers} illustrates this feature by comparing 9 points sampled from the functions $f(x)$ specified in the plots to the interpolation using the B\'ezier curve ${\cal B}^{(n)}(x)$ constructed according to Eqs.~(\ref{eq:CMTPsquare}) and (\ref{eq:CMTPrectangular}). We also show truncated monomial approximations for this curve, $\sum_{l=0}^N \bar{c}_l \ (1-x)^l$ with $0\leq N <n$. The input PDFs satisfy $f(x=1)=0$. We therefore expect the $N=0$ truncation to vanish if the B\'ezier interpolation is constructed properly, signifying that $c_n=\bar{c}_0=0$. Indeed, the $N=0$ truncation is consistent with zero in all subfigures of Fig.~\ref{fig:Beziers}.

Figure~\ref{fig:Beziers}(a) shows ${\cal B}^{(n)}(x)$ for an input function given by a fourth-degree polynomial, $f(x)=30\, x^2\, (1-x)^2$. The red points denote $\{x_i,p_i=f(x_i)\}$ given by the exact input function. The interpolating polynomial is readily found as ${\cal B}^{(4)}(x)$ using Eq.~(\ref{eq:CMTPrectangular}), or equivalently as ${\cal B}^{(n)}(x)$ with $c_l=0$ for $n, \ l>4$. The monomial term with a linear $1-x$ dependence, denoted as $N=1$, vanishes in this simple case.

For a more elaborate input function assumed in Figs.~\ref{fig:Beziers}(b,c,d), 
\begin{equation}
f(x)=200\, x^2\, (1-x)^2\, \left(1 - 2.9\sqrt{x(1-x)} + 2.3 x\, (1-x)\right),
\label{f200}
\end{equation}
we must resort to ${\cal B}^{(8)}(x)$ with $n=k=8$ to obtain interpolation that goes through all points $p_i$ according to Eq.~(\ref{eq:CMTPsquare}). One can see from the figures that linear terms ($N=1$) are present in the respective monomial expansions, even though the input function does not contain linear terms. The slopes of the $N=1$ terms are different in the three panels, the monomial expansion converges too slowly for $x<0.8$. The coefficients of the linear and higher-$l$ terms depend on the ranges and spacings of $x_i$, which are varied in Figs.~\ref{fig:Beziers}(b,c,d). The ambiguity in the $N=1$ term is introduced by the correlation of the  coefficient $\bar{c}_1$ with the coefficients $\bar{c}_l$ with high $l$. To pin down the linear term, we must restrict the fit to the highest portion of the $x$ range, such as $x>0.9$. It appears that finding the low-$l$ monomial terms with good accuracy requires one to fit precise data in the interval that extends closely to the end point $x=1$, where higher-twist effects are likely important. 

With typical input PDF shapes and fewer than about 10 input points, we find that interpolation by B\'ezier curves is numerically stable over the range $x_0 \leq x \leq x_k$: the B\'ezier curve goes through all input points and may differ from the input function in high-order terms between the points. Outside of the $x$ range covered by the data points, extrapolation can be unstable.

\subsection{Effective large-$x$ exponent\label{sec:A2eff}}
The example in Fig.~\ref{fig:Beziers} demonstrates that {\it mimicry} of the fitted functional forms impedes determination of the lowest powers in the monomial $(1-x)$ expansion even in an idealized fit to a few "data" points without uncertainties. An interpolation or fit by a high-degree polynomial may render terms with low powers of $(1-x)$ that are not present in the fitted function, and which depend on how the data are sampled. In QCD, there is no reason to expect that coefficients $\bar{c}_l$ for high powers of $(1-x)$ are suppressed in the PDFs. Statistical and systematic errors in the measurements also get in the way of the determination of the analytic $(1-x)$ dependence.   

The remainder of the article will follow a less pretentious path. Predictions of QCR's and various nonperturbative models suggest that, in the $x_{(B)}\to 1$ limit, the structure functions or PDFs, denoted collectively as ${\cal F}(x_{(B)}, Q^2)$, behave as 
\begin{equation}
\label{eq:calF}
{\cal F}(x_{(B)}, Q^2)=(1-x_{(B)})^{A_{2}}\times \Phi(1-x_{(B)})
\;,
\end{equation}
where $\Phi(1-x_{(B)})$ is a slowly varied function. Based on this observation, it is natural to define 
\begin{equation}
\label{eq:A2Def}
A_2^{\mbox{\tiny eff}}\left[{\cal F}(x_{(B)},Q^2)\right]\equiv \frac{\partial \ln\left({\cal F}(x_{(B)}, Q^2)\right)}{\partial \ln\left(1-x_{(B)}\right)}
\;,
\end{equation}
with the expectation that $A_2^{\mbox{\tiny eff}} \approx A_2$ when the logarithmic derivative of $\Phi(1-x_{(B)})$ is small.\footnote{   
A similar definition was introduced in Ref.~\cite{Ball:2016spl}.} We will compare theoretical predictions for $A_2$ presented in Sec.~\ref{sec:QCR} with the $A_2^{\mbox{\tiny eff}}\left[{\cal F}(x_{(B}),Q^2)\right]$ values obtained from a phenomenological PDF ensemble. 

\subsection{The CT18 PDF ensemble and estimation of PDF uncertainty \label{sec:CT18}}
In the present analysis, we focus on the results obtained with the CT18 global QCD  analysis~\cite{Hou:2019efy}. The CT18 PDFs are determined by fitting NNLO theoretical cross sections to 40 experimental data sets with a total of 3681 data points. The fitted scattering processes -- DIS, production of lepton pairs, jets, and $t\bar t$ pairs -- cover a large kinematic region that extends up to $x=0.75$ in the typical momentum fraction and down to $Q=2$ GeV in the factorization scale. 

The CT18 functional form is given by
\begin{equation}
\label{eq:CT18FF}
f_{a/A}(x, Q_0^2)=x^{A_{1, a}}(1-x)^{A_{2, a}}\times \Phi_a(x).
\end{equation}
Functions $\Phi_a(x)$ are parametrized by B\'ezier curves ${\cal B}^{(n)}(y)$ with $n=4-5$ and $y\equiv\sqrt{x}$ or a similar scaling function; see Appendix C in \cite{Hou:2019efy}. At the initial scale of the fit, $Q_0=1.3$ GeV, the functional forms provide the initial condition for DGLAP equations that predict PDFs at an arbitrary $Q>Q_0$. Each interpolating polynomial $\Phi_a(x)$ reduces to a non-zero constant at $x\to 0$ or $1$, so that the exponents $A_{1,a}$ and $A_{2,a}$ control the behavior of the individual PDFs when approaching these limits. While in principle all $A_{2,a}$ parameters can be determined from the data, in practice not all their combinations result in non-negative PDFs or physically acceptable values of flavor-dependent observables. The CT18 fit imposes a requirement $A_{2,u_{\mbox{\tiny V}}}=A_{2,d_{\mbox{\tiny V}}}$ to guarantee a finite value of $d(x,Q^2)/u(x,Q^2)$ at $x\rightarrow 1$ or, equivalently, a nontrivial asymptotic value of $F_2^p(x,Q^2)/F_2^n(x,Q^2)$. We thus expect 
\begin{equation}
\lim_{x\rightarrow 1} \frac{A_{2,d_{\mbox{\tiny V}}}^{\mbox{\tiny eff}}(x)}{A_{2,u_{\mbox{\tiny V}}}^{\mbox{\tiny eff}}(x)} = \frac{A_{2,d_{\mbox{\tiny V}}}}{A_{2,u_{\mbox{\tiny V}}}} = 1,
\label{eq:A2d2A2u}
\end{equation}
where the limit is reached only at very high $x$ values that are outside of the $x$ region covered by the experimental measurements.

On the other hand, at $x$ values below 0.8, where sufficient amount of experimental data exists, the polynomials $\Phi_a(x)$ with $a=u$ or $d$ are flexible enough to allow a variety of functional behaviors of the $u$ and $d$ PDFs. Therefore, at $x < 0.8$, $A_{2,d_{\mbox{\tiny V}}}^{\mbox{\tiny eff}}(x)$ can be quite different from $A_{2,u_{\mbox{\tiny V}}}^{\mbox{\tiny eff}}(x)$.
\\

The CT18 ensemble consists of the central PDF set and 58 error PDF sets that can be used to estimate the PDF uncertainty in $A_2^{\mbox{\tiny eff}}$ according to the master formulas of Refs.~\cite{Nadolsky:2001yg,Lai:2010vv}. A variety of sources contribute to this PDF uncertainty, including experimental, theoretical, parametrization, and methodological uncertainties \cite{Kovarik:2019xvh}. Among these, the uncertainty introduced by the choice of the PDF functional forms
has been examined in the CT18 study by examining the spread of the PDFs in 250 candidate fits with 
alternative functional forms or methodological settings, as explained in Sec.~III.C.3 of \cite{Hou:2019efy}. The tolerance on the nominal Hessian error sets has been selected so as to cover, on average, the spread of the best-fit values in the alternative fits. Thus, the CT18 Hessian uncertainty covers the spread of results with the alternative parametrization choices, with the exception of the extrapolated $x$ regions, $x > 0.7$, where some of the explored best fits fall outside of the nominal CT18 error band. We therefore quote the uncertainty on $A_2^{\mbox{\tiny eff}}$ as the envelope constructed from the CT18 Hessian uncertainty at the 68\% probability level and the extreme variations of $A_2^{\mbox{\tiny eff}}$ obtained with the extended set of 363 alternative functional forms. 

\begin{figure}[tb]
\centering
\includegraphics[width=.5\textwidth]{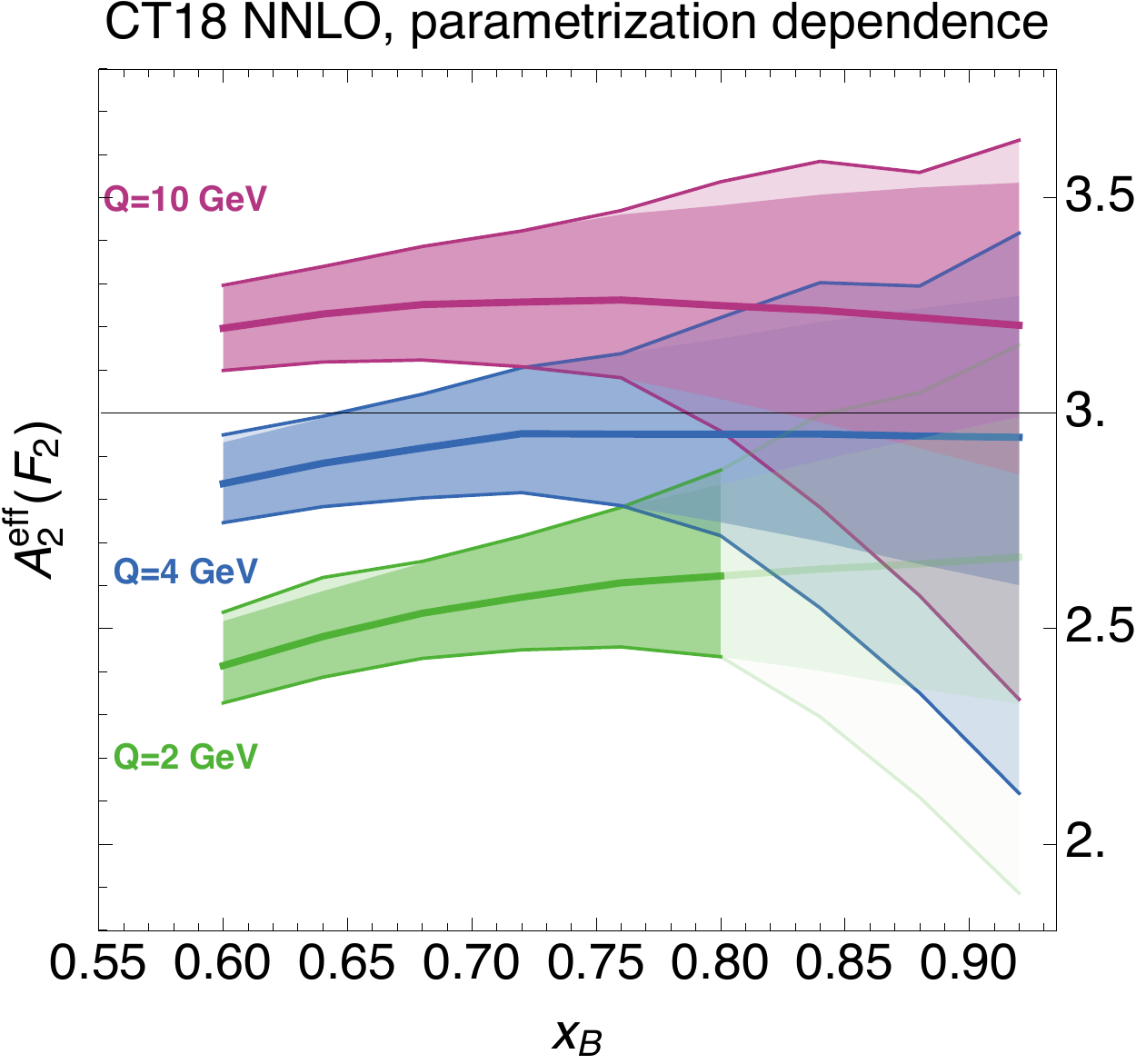}
\caption{The effective exponent $A_2^{\mbox{\tiny eff}}$ for the structure function $F_2(x_{\rm B},Q^2)$ as a function of $x_{\rm B}$ and for the $Q$ values of $2,\, 4$ and $10$ GeV represented in green, blue and magenta, respectively. The central curve of each error band represents the CT18 NNLO central value, the dark shaded band is the asymmetric Hessian error \cite{Nadolsky:2001yg,Lai:2010vv} at the 68\% probability level. The extreme curves correspond to the envelope of the Hessian and parametrization uncertainties estimated as in Sec.~\ref{sec:CT18}. The transparent part of the $Q=2$ GeV band corresponds to the region with $W^2<2m_p^2$, approximately corresponding to the resonance region in DIS. The reference prediction from the QCRs is shown by a line at $A_2^{\mbox{\tiny eff}}(F_2)=3$.}
\label{fig:OverTheRainbow}
\end{figure}

\subsection{An effective exponent for a DIS structure function\label{sec:a2effF2p}}
In deep inelastic scattering, a structure function $F(x_B,Q^2)$ of the proton can be predicted in terms of {\it phenomenological} PDFs $f_{a/p}(x,\mu^2)$ as in the QCD factorization formula in Eq.~(\ref{eq:DISfactorization}).  Figure~\ref{fig:OverTheRainbow} shows  the effective $(1-x_{\rm B})$-power $A_2^{\mbox{\tiny eff}}$ for $F_2(x_B,Q^2)$ computed according to Eq.~(\ref{eq:A2Def}) with the CT18 NNLO set~\cite{Hou:2019efy} for three values of $Q^2$. These are predictions for the leading-twist contribution to $F_2(x_B,Q^2)$ that reflect a combination of constraints from 40 diverse experiments, and which may not be directly comparable to the actual DIS data because of the limitations discussed in Sec.~\ref{sec:LargeXRealQCD}. At $Q=2$ GeV, the interval $x_B \gtrsim 0.8$ corresponds to DIS in the resonant region, see Sec.~\ref{sec:LargeXRealQCD}, where the smooth behavior of $F_2(x_B,Q^2)$ that we predict is modified by complex local features that do not obey the QCR's. We indicate the $x_B$ values lying in the resonant DIS region, approximately corresponding to $W^2 < 2 m_p^2$, by using the semi-transparent fill for a part of the error band for $Q=2$ GeV. 

Each error band in Fig.~\ref{fig:OverTheRainbow} consists of an inner (darker) part, indicating the 68\% probability level Hessian uncertainty, and the outer (lighter) part, representing the envelope formed by the Hessian uncertainty and the exponents~(\ref{eq:A2Def}) for the alternative functional forms, as explained in Sec.~\ref{sec:CT18}. We see that the magnitudes of the Hessian and envelope uncertainties are similar for the region where data are available, {\it i.e.} $x < 0.75$. On the other hand, in the region in which the PDFs are unconstrained by the data, the error reflecting the choice of parametrization dominates. The envelope indicated by the outer error band represents a more conservative estimate of the uncertainty at each $Q^2$.

We now compare the $A_2^{\mbox{\tiny eff}}\left[{\cal F}(x_{B},Q^2)\right]$ from the CT18 fit to the prediction $A_2=3$ from the QCR's for $F_2(x_B,Q^2)$ in a proton given by Eq.~(\ref{eq:QCR_F2p}). The $A_2^{\mbox{\tiny eff}}$ values for $Q=2$ GeV in Fig.~\ref{fig:OverTheRainbow} clearly fall below the QCR prediction. On the other hand, starting from a scale of about $4$ GeV, the expected exponent of three is attained at the highest $x$ within the error bands. 

The $A_2^{\mbox{\tiny eff}}$ values in Fig.~\ref{fig:OverTheRainbow} substantially depend on $x_B$ and $Q^2$. The true limit of the QCR's would imply weak dependence on either $x_B$ or $Q^2$, indicating that neither multiparticle final Fock states nor anomalous dimensions are important. In accordance with the discussion in Sec.~\ref{sec:QCR}, the QCR's would be realized when the Feynman diagrams in Fig.~\ref{fig:QCRdiagrams} dominate. 

As we alluded in Sec.~\ref{sec:LargeXRealQCD}, we do not expect the conditions for the QCR's to be fully met in a global PDF fit. Various simplifications in the fitted processes limit the anticipated accuracy, for example, due to the neglect of process-dependent power-suppressed terms. Nevertheless, Fig.~\ref{fig:OverTheRainbow} shows that the CT18 values for $A_2^{\mbox{\tiny eff}}\left[{\cal F}(x_{B},Q^2)\right]$ are consistent with the QCR prediction of three within about one unit. 
Keep in mind that the bound-state wave function for the hadronic target predicts parton distributions in a free hadron.
On the other hand, the nucleons --and {\it a fortiori}  the pions-- probed in the global fits are not truly free at some level: their partons feel the other hadrons present in the scattering event before or after the hard scattering. At the leading power, the effect of the soft QCD background field created by spectator hadrons on the propagation of partons entering the scattering is given by the Wilson line in the $\overline{MS}$ PDF in Eq.~(\ref{eq:MSbarPDF}). Additional radiation of this kind and power-suppressed terms may modify the $x$ dependence compared to the QCR prediction. 

    \subsection{Effective exponents for PDFs\label{sec:a2effPDF}}
\begin{figure}[tb]
\centering
\includegraphics[width=.35\textwidth]{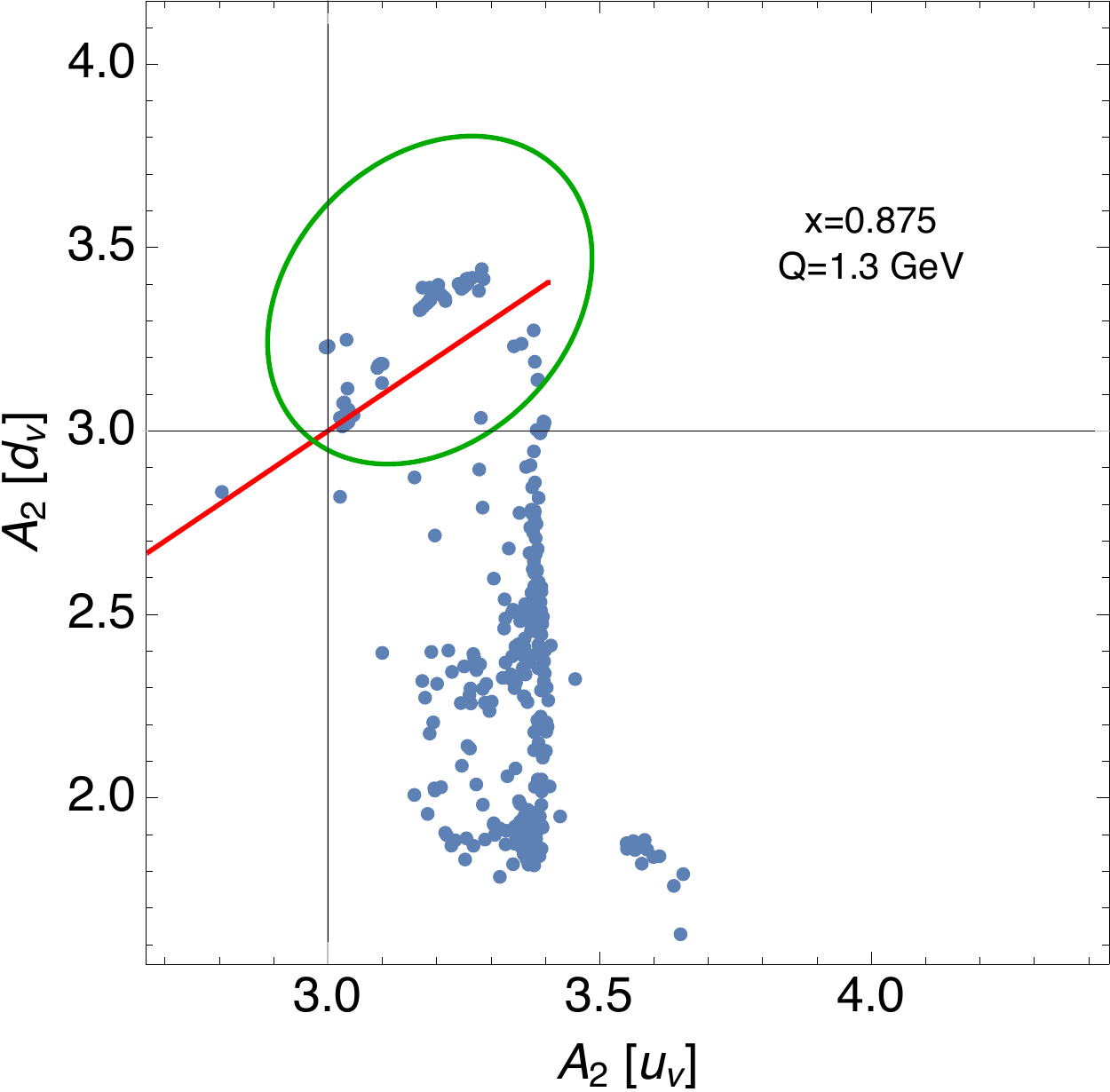}
\includegraphics[width=.35\textwidth]{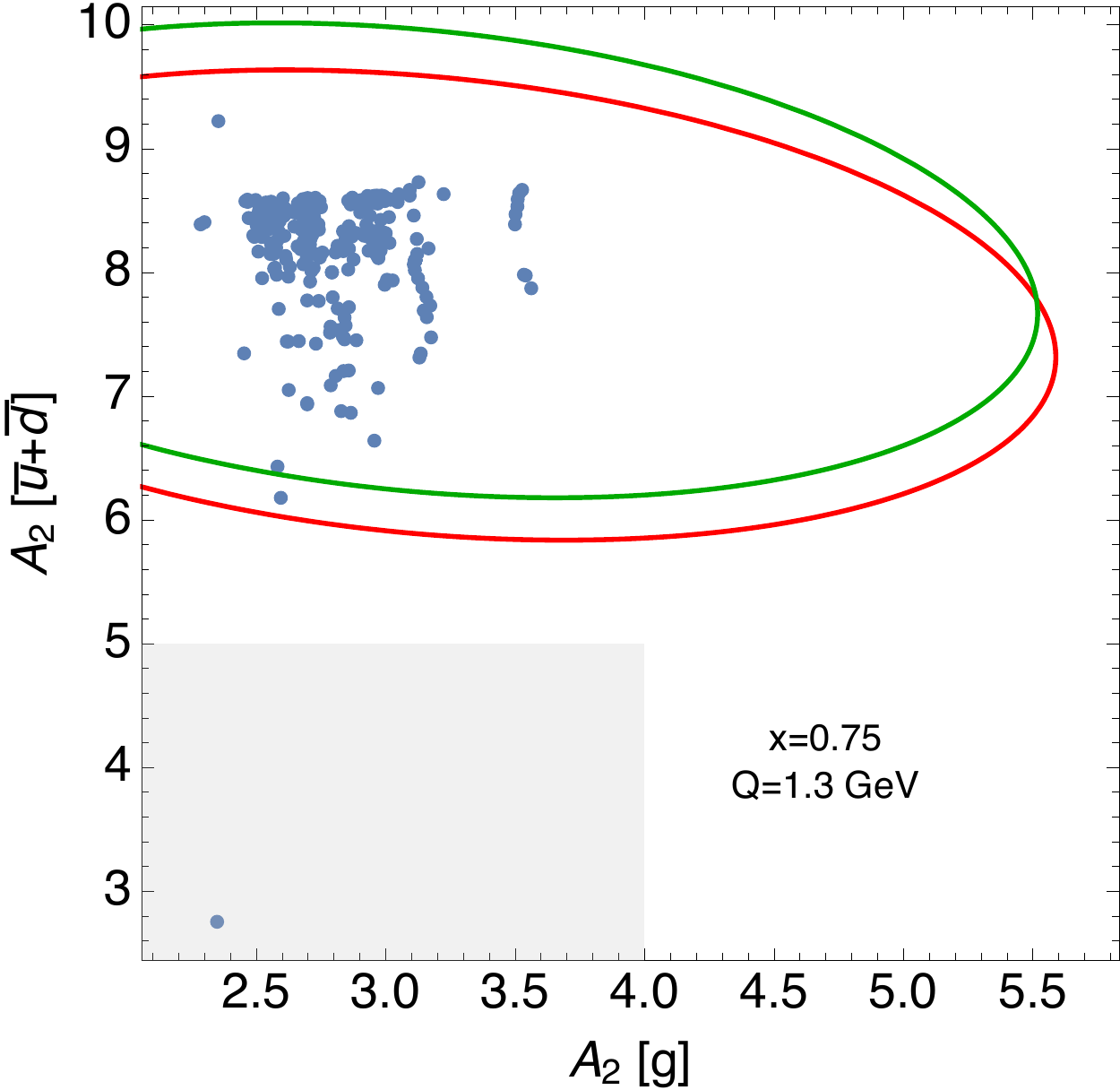}
\caption{The effective exponent $A_2^{\mbox{\tiny eff}}$ for 363 alternative parametrizations of CT18 NNLO (blue points), compared  to the CT18 NNLO Hessian error ellipse at $68\%$ c.l. for  $A_2^{\mbox{\tiny eff}}[\mbox{\scriptsize CT18NNLO}]$ (green ellipse) and the tabulated $A_2$ parameters of the CT18NNLO error ensemble at $68\%$ c.l. (red line or red ellipse). Both plots are shown at $Q_0=1.3$ GeV. The left panel is for the PDF flavors $u_{\mbox{\tiny V}}$ vs.  $d_{\mbox{\tiny V}}$ at $x=0.875$. The right panel is for $g$ vs.  ${\bar u}+{\bar d}$  at $x=0.75$. The lines in the left subfigure show the expected values. The gray rectangle in the right subfigure shows the region forbidden by the QCRs, see text.}
\label{fig:CT18nn_A2u_A2d_ori}
\end{figure}
\begin{figure}[tb]
\centering
\includegraphics[width=.75\textwidth]{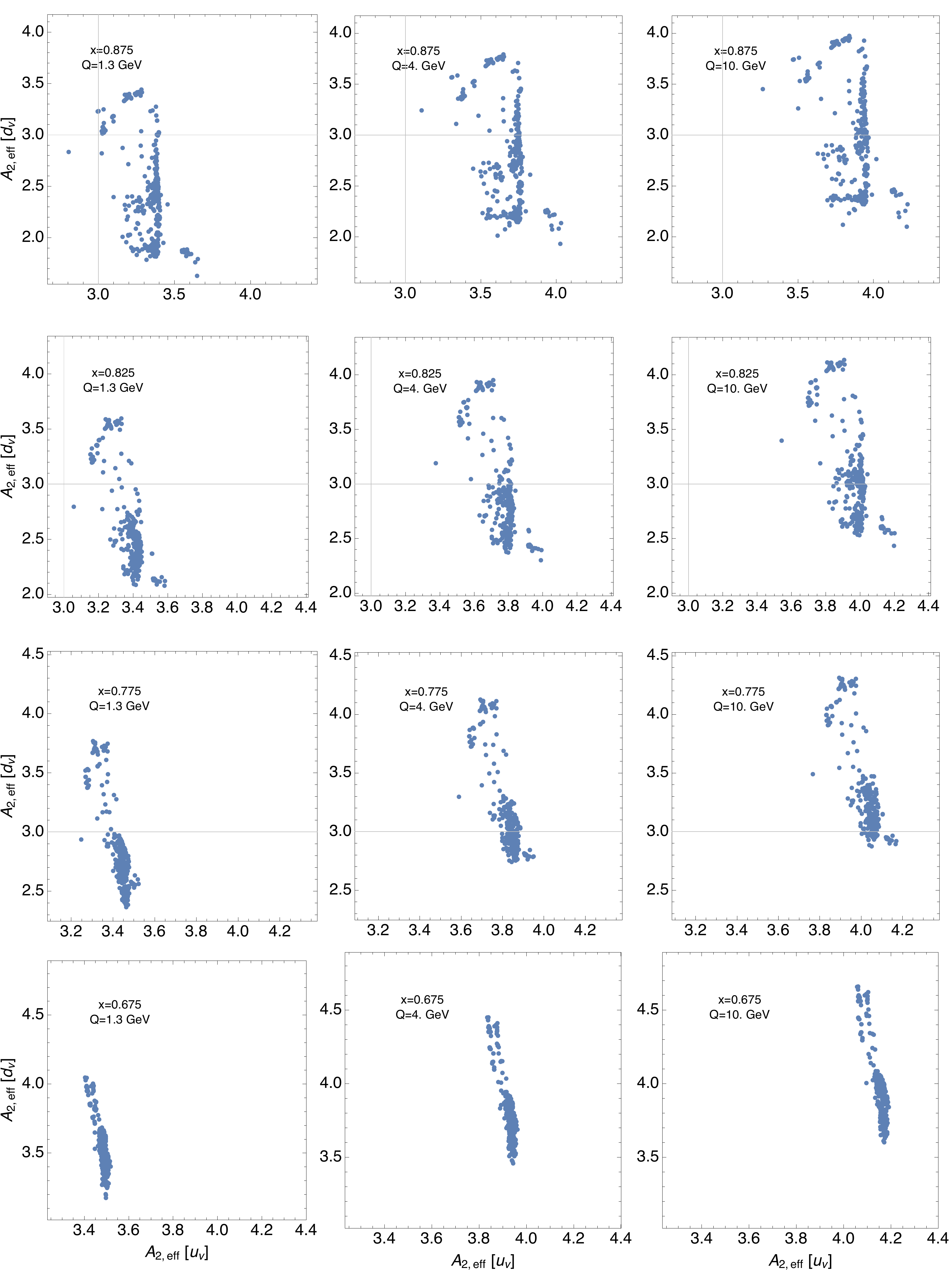}
\caption{The effective exponent $A_2^{\mbox{\tiny eff}}$ for the alternative parametrizations for the PDF flavors $u_{\mbox{\tiny V}}$ vs.  $d_{\mbox{\tiny V}}$. The $x$ values, from the upper to lower row, are $0.875, 0.825, 0.775, 0.675$. The $Q$ values, from left to right, are $1.3, \, 4$, and $10$ GeV. The lines $A_2^{\mbox{\tiny eff}}(u_{\mbox{\tiny V}},d_{\mbox{\tiny V}})=3$ are shown for reference.}
\label{fig:CT18nn_A2u_A2d}
\end{figure}

\begin{figure}[tb]
\centering
\includegraphics[width=.75\textwidth]{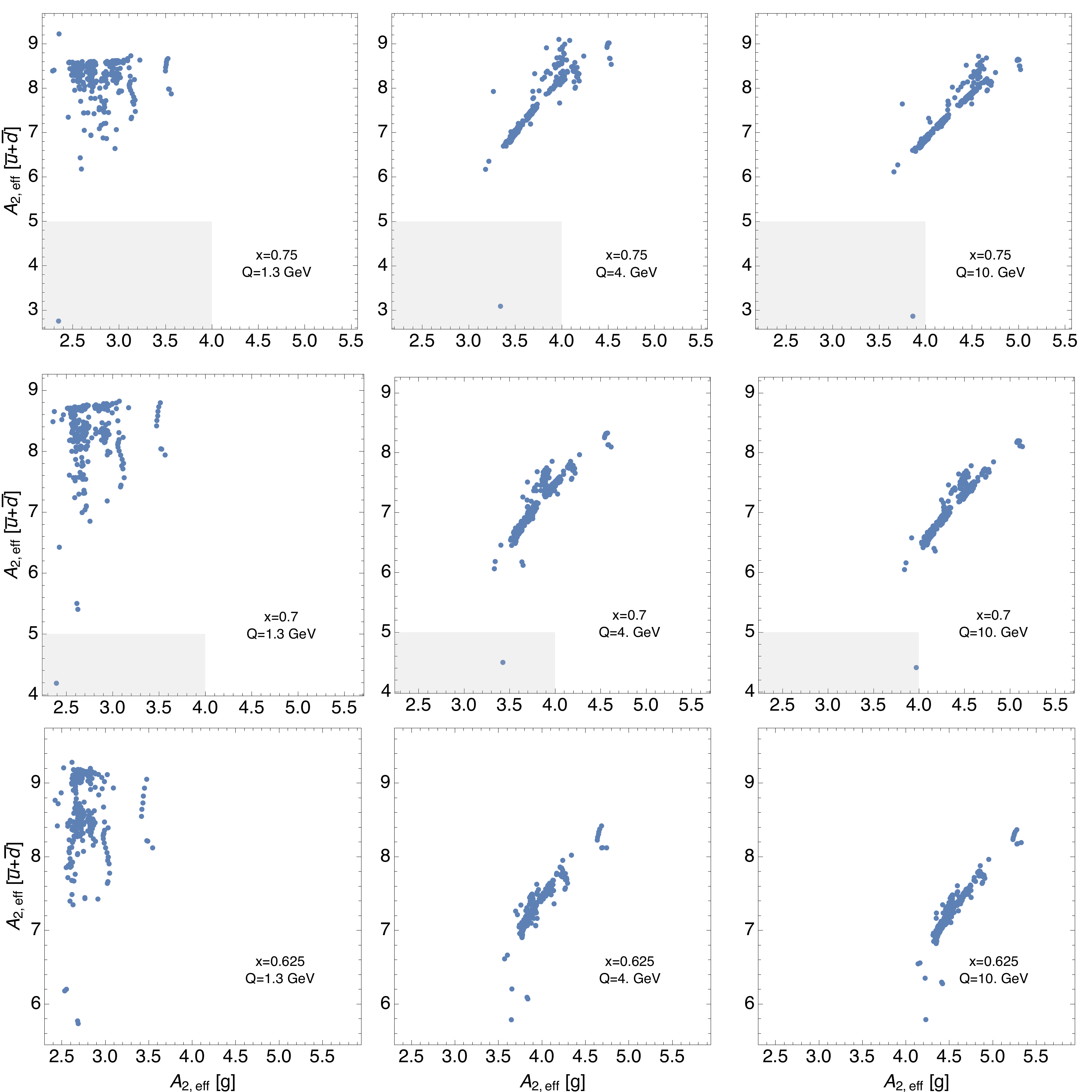}
\caption{The effective exponent $A_2^{\mbox{\tiny eff}}$ for the PDF flavors $g$ vs.  $\bar{u}+\bar{d}$. The $x$ values, from the upper to lower row, are $0.875, 0.825, 0.775, 0.675$. The $Q$ values, from left to right, $1.3, \, 4$, and $10$ GeV. The rectangle shows the region  $A_2^{\mbox{\tiny eff}}(g)< 4$ and $A_2^{\mbox{\tiny eff}}(\bar{u}+\bar{d})< 5$ that is disfavored by the QCRs.}
\label{fig:CT18nn_A2g_A2ubdd}
\end{figure}

We will now investigate effective exponents $A_2^{\mbox{\tiny eff}}$  for individual PDFs. 
In the CT18 global fit, {\it universal} PDFs enter theoretical cross sections for the fitted processes as shown in  Eqs.~(\ref{eq:DISfactorization}) and (\ref{eq:DYfactorization}) for DIS and DY. The hadronic cross sections are evaluated up to NNLO in $\alpha_s$. Equation~(\ref{eq:QCR_A2iQ2}) states the QCR predictions for the $A_2$ exponents for PDFs of various flavors. These predictions can be compared with $A_2^{\mbox{\tiny eff}}$ computed for the respective phenomenological PDFs.

First, we plot, in the left subfigure of Fig.~\ref{fig:CT18nn_A2u_A2d_ori}, the nominal parameters $A_{2,u_{\mbox{\tiny V}}}$ and $A_{2,d_{\mbox{\tiny V}}}$ in the parametrizations for $u$ and $d$ valence quarks, introduced as in Eq.~(\ref{eq:CT18FF}). As summarized in Sec.~\ref{sec:CT18}, the CT18 fit assumes these parameters to be the same. The figure shows the red line that corresponds to the best-fit CT18 value for $A_{2,u_{\mbox{\tiny V}}}=A_{2,d_{\mbox{\tiny V}}}$ and its 68\% c.l. Hessian uncertainty. 

In the same subfigure, the green ellipse shows the 68\% c.l. region for the {\it effective} exponents $A_2^{\mbox{\tiny eff}}$ computed using the CT18 Hessian error PDF set at $Q=1.3$ GeV and $x=0.875$ according to Eq.~(\ref{eq:A2Def}).
Finally, blue scattered points in the left subfigure are for the $A_2^{\mbox{\tiny eff}}$ combinations obtained with 363 alternative parametrizations of CT18 PDFs. The lines indicate the QCR prediction of three for each exponent. 

We see in the left Fig.~\ref{fig:CT18nn_A2u_A2d_ori} that, at this high value of $x$ and the initial scale $Q_0$, the nominal parameters $A_{2,u_{\mbox{\tiny V}}}$ and $A_{2,d_{\mbox{\tiny V}}}$ are consistent with the QCR predictions within the PDF uncertainty. While the nominal $A_{2,u_{\mbox{\tiny V}}}$ and $A_{2,d_{\mbox{\tiny V}}}$ are set to be equal in the initial functional forms, the effective coefficients for $u_{\mbox{\tiny V}}$ and $d_{\mbox{\tiny V}}$ at $x=0.875$ turn out to be slightly different. The distribution of the scatter points is narrower for the up valence than for the down valence, implying that the PDF for $d_{\mbox{\tiny V}}$ is less constrained by the data at large $x$.

Fig.~\ref{fig:CT18nn_A2u_A2d} shows how $A_2^{\mbox{\tiny eff}}$ for $u_{\mbox{\tiny V}}$ and $d_{\mbox{\tiny V}}$ obtained with the alternative parametrizations change when $Q$ takes the values of 1.3, 4, and 10 GeV, and $x$ varies between 0.675 and 0.875.
The left column, corresponding to $Q=Q_0=1.3$ GeV, illustrates the $x$ dependence of $A_2^{\mbox{\tiny eff}}$ at the initial scale of DGLAP evolution. The scattered clusters for $u_{\mbox{\tiny V}}$ become narrower toward smaller $x$ values -- going down from the top to the bottom row. Going from the left to the right in each row, we observe the effect of DGLAP evolution when increasing $Q$. The shapes of the point distributions are largely preserved when increasing $Q$, while the distributions shift toward higher $A_2^{\mbox{\tiny eff}}$ as a whole. The {\it effective} $A_2^{\mbox{\tiny eff}}\left(u_{\mbox{\tiny V}}(x,Q)\right)$ is larger than the expected value of three for all the scales considered. On the other hand, $A_2^{\mbox{\tiny eff}}\left(d_{\mbox{\tiny V}}(x,Q)\right)$ is as low as two for some parametric forms.  Overall, the figure demonstrates non-negligible dependence of $A_2^{\mbox{\tiny eff}}$ on $x$ (possibly caused by more Fock states contributing to scattering at smaller $x$) and on $Q$ (reflecting the anomalous dimensions for the $A_2$ exponents). 

Similar plots for $g$ and $\bar{u}+\bar{d}$ PDFs are shown in the right Fig.~\ref{fig:CT18nn_A2u_A2d_ori} for the 
$A_2^{\mbox{\tiny eff}}$ values at $Q_0=1.3$ GeV and $x=0.75$, and in Fig.~\ref{fig:CT18nn_A2g_A2ubdd} for the $x$ and $Q$ dependence. These PDFs quickly vanish at very large $x$, thus we limit the respective comparisons to the region $x\leq 0.75$ to avoid numerical issues. 

Since the nominal $A_2$ parameters are not constrained to be the same for $g$ and $\bar{u}+\bar{d}$, the 68\% c.l. Hessian uncertainty region in the right Fig.~\ref{fig:CT18nn_A2u_A2d_ori} is given by an ellipse and not by a line. The Hessian uncertainty region (red ellipse) on $A_2^{\mbox{\tiny eff}}$ values for these flavors agrees well with the respective nominal $A_2$ values (green ellipse), as well as with the $A_2^{\mbox{\tiny eff}}$ obtained with alternative parametrizations. 

According to Fig.~\ref{fig:CT18nn_A2g_A2ubdd}, the $A_2^{\mbox{\tiny eff}}$ values for $g$ and $\bar{u}+\bar{d}$ depend weakly on the $x$ value, taken to be $x=0.625$, 0.7, and 0.75. The $Q$ dependence is significant for the gluon $A_2^{\mbox{\tiny eff}}$ and much weaker for $\bar{u}+\bar{d}$. At $Q_0=1.3$ GeV, neither $A_2^{\mbox{\tiny eff}}$ is truly compatible with the respective QCR predictions: the gluon $A_2^{\mbox{\tiny eff}}$ of 2-4 is systematically lower than the QCR prediction of 4, while the sea-quark $A_2^{\mbox{\tiny eff}}$ of 6-9 tends to be higher than 5. 

The $Q$ dependence due to the singlet DGLAP evolution is very pronounced for both types of PDFs. The gluon 
{\it effective} exponent grows above four starting from about $4$ GeV, now in compliance with the QCRs. 
The $A_2^{\mbox{\tiny eff}}$ for the sea combination $\bar{u}+\bar{d}$ has a large uncertainty at $Q_0$ but quickly correlates with the gluon once the DGLAP evolution is turned on.

We can compare the effective exponents for CT18 NNLO PDFs with those computed for the MMHT14 and NNPDF3.0 sets. We find reasonable agreement between the effective exponents obtained based on these three PDF ensembles. We also broadly agree with the observations presented in Ref.~\cite{Ball:2016spl} for a low $Q$ scale.\footnote{Let us remark that the NNPDF3.0 results must be averaged, as mentioned in the cited reference. }

\begin{figure}[p]
\centering
\vspace{0.15cm}
\includegraphics[height=.4\textheight]{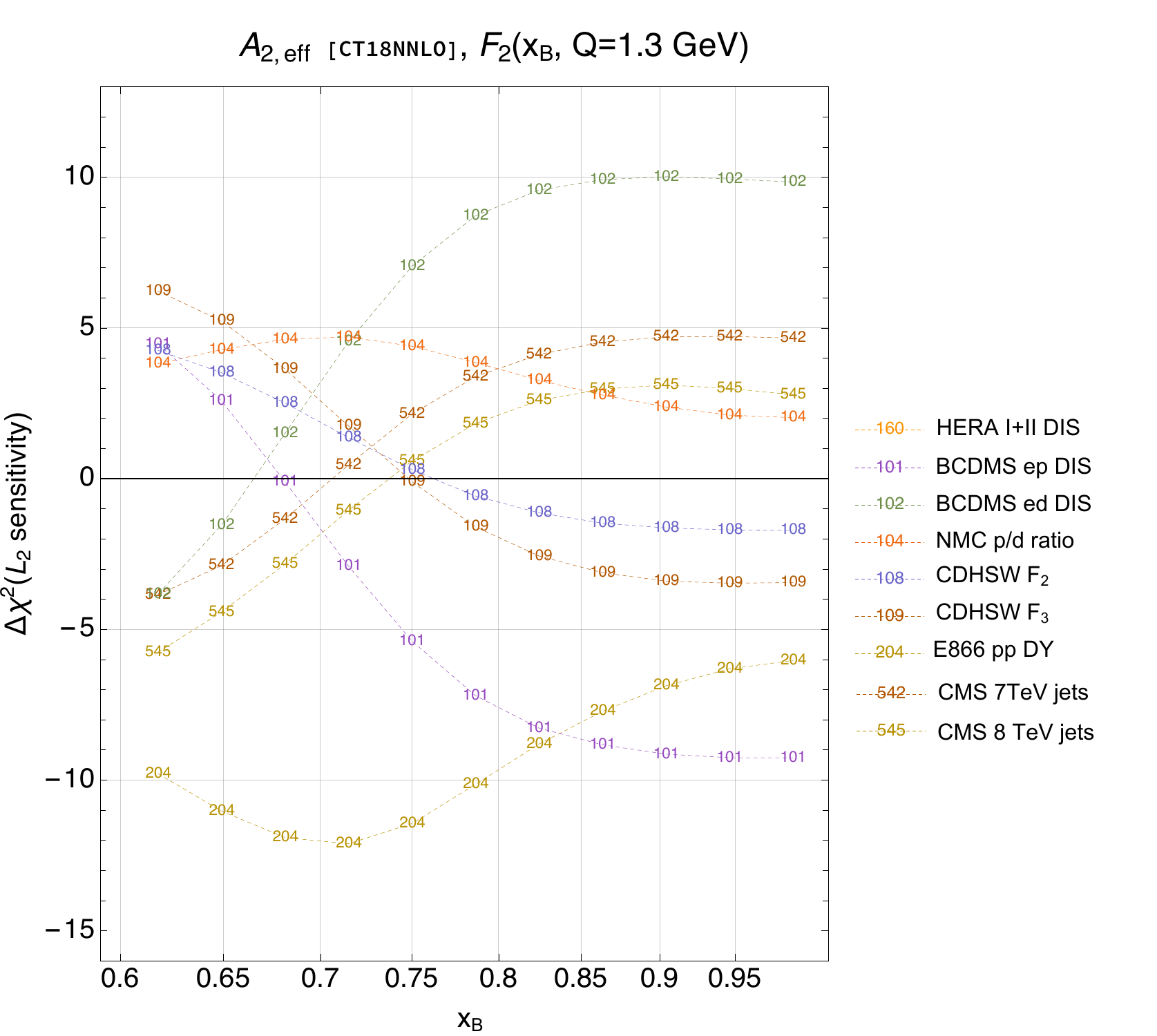}\\
\includegraphics[height=.4\textheight]{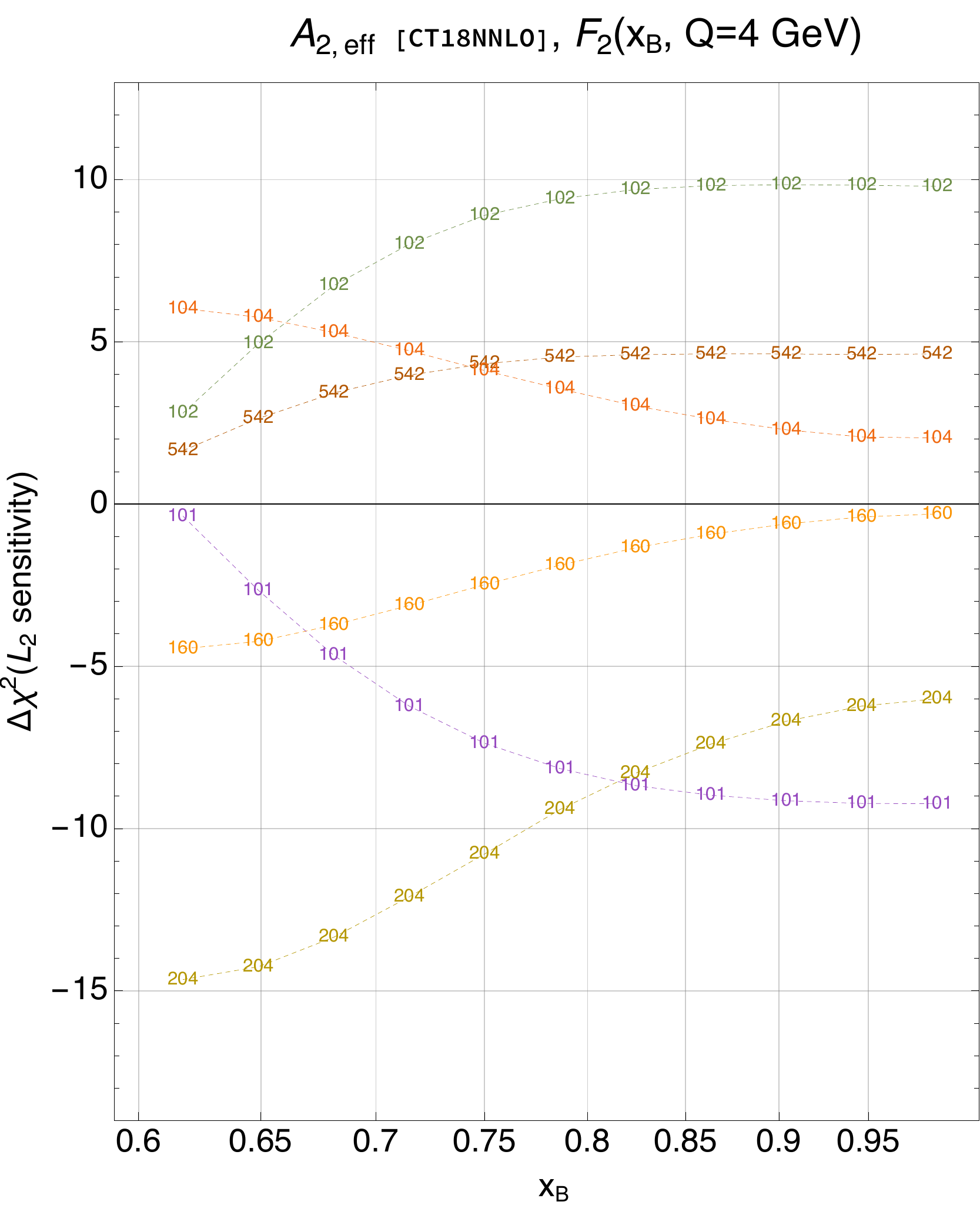}
\includegraphics[height=.4\textheight]{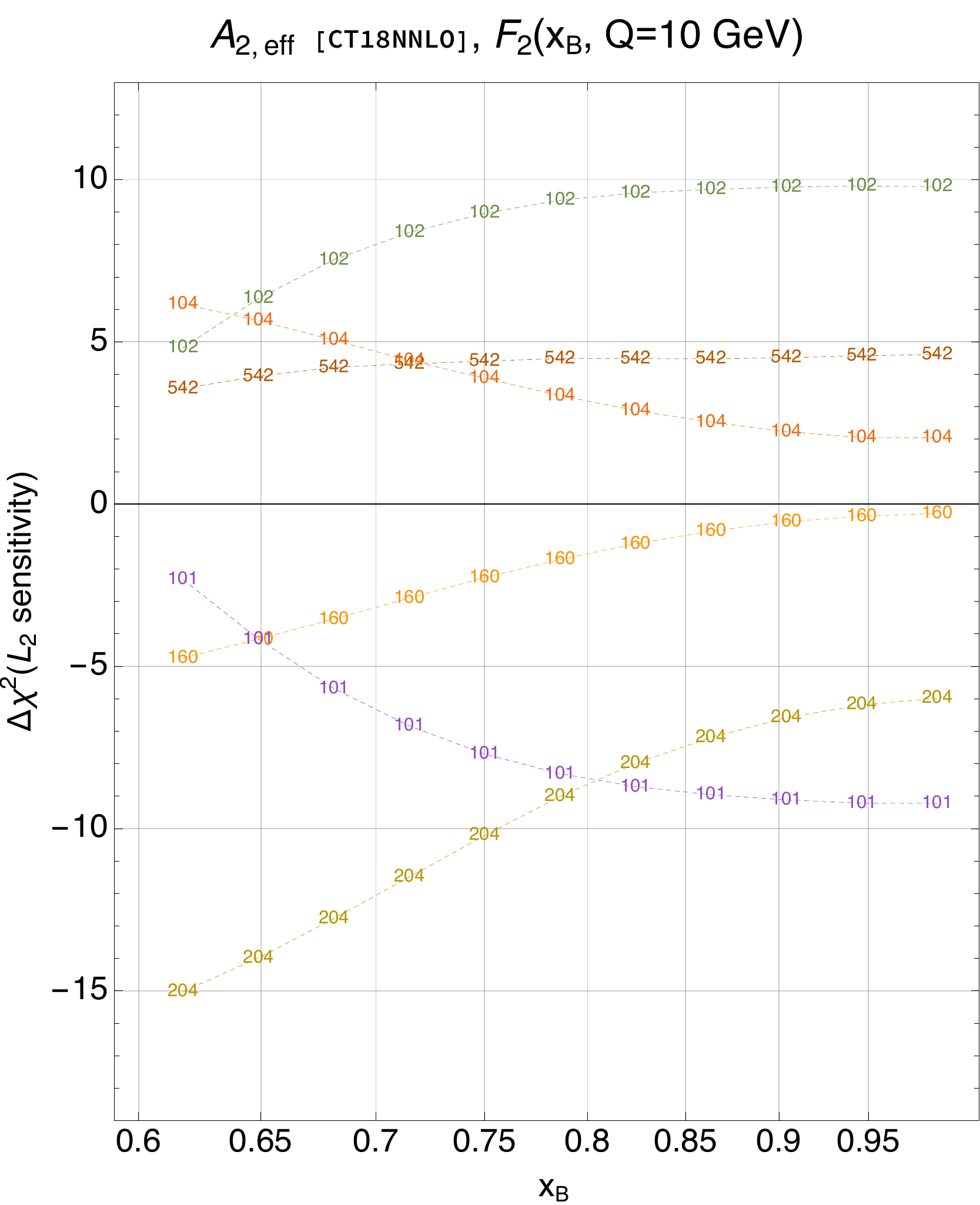}
\caption{ $L_2$ sensitivity to the effective exponent $A_2^{\mbox{\tiny eff}}$ for the structure function $F_2(x_B,Q^2)$ evaluated with the CT18NNLO PDF ensemble, as a function of $x_B$ for the $Q$ values of $1.3,\, 4$ and $10$ GeV.}
\label{fig:OTRSensitivity}
\end{figure}

We learn several things from these comparisons. The phenomenological effective exponents and QCR predictions better agree for the valence $u$ and $d$ quarks. The agreement is not so good for the gluon and especially the sea quark PDFs. 

The assumptions justifying the quark counting rules hold when the scale dependence is small.  In reality, Figs.~\ref{fig:CT18nn_A2u_A2d} and \ref{fig:CT18nn_A2g_A2ubdd} demonstrate pronounced scale dependence, 
suggesting that higher-order QCD radiation, producing multi-parton final states, is not entirely negligible in the fitted processes. 

Focusing on DIS for a moment, the numerical effect of the multi-parton final states at various $Q^2$ and $W^2$ is twofold. When $Q^2$ is increased at a fixed $x$ value, QCD radiation, evaluated in the logarithmic approximation by DGLAP equations, increases the effective power $A_2^{\mbox{\tiny eff}}$ for valence quarks and gluons --- see \cite{Ball:2016spl} and references therein. The growth of the corresponding $A_2^{\mbox{\tiny eff}}$ in our figures is consistent with this expectation, while the $Q^2$ dependence for antiquark $A_2^{\mbox{\tiny eff}}$ is less clear. 

When $W^2$ is increased (and $x_{\rm B}$ is decreased) at a fixed $Q^2$, new final states may be produced and introduce terms with high powers of $1-x$ in the PDF expressions. As discussed in Sec.~\ref{sec:Mimicry}, such terms can {\it increase} or {\it reduce} the effective leading power $A_2^{\mbox{\tiny eff}}$, as compared to the QCR prediction, and introduce dependence of $A_2^{\mbox{\tiny eff}}$ on $x_{(B)}$.

    \subsection{Process dependence of effective exponents\label{sec:a2effByProcess}}
In Secs.~\ref{sec:a2effF2p} and \ref{sec:a2effPDF}, we presented the effective exponents $A_2^{\mbox{\tiny eff}}$ and their  uncertainties for leading-power DIS structure functions and PDFs determined based on the totality of 40 experiments fitted in the CT18 global analysis. We will now examine the agreement of individual experimental data sets in their preferences for the large-$x$ behavior of PDFs quantified by
$A_2^{\mbox{\tiny eff}}$. Toward this goal, we will employ a statistical indicator called the $L_2$ sensitivity \cite{Hobbs:2019gob}, following its applications in the CT18 analysis \cite{Hou:2019efy} to examine agreement between the experimental data sets. The $L_2$ sensitivity is constructed from the values $A_2^{\mbox{\tiny eff}}$ and  goodness-of-fit function $\chi^2_E$ for each fitted experiment $E$, computed for each Hessian eigenvector set of the CT18 NNLO ensemble. See the relevant equations in Ref.~\cite{Hobbs:2019gob}. 

Figure~\ref{fig:OTRSensitivity} plots the $L_2$ sensitivity of several fitted experiments to the effective $A_2^{\mbox{\tiny eff}}$ for the proton structure function, $F_2^p(x_B,Q^2)$, evaluated at the $x_B$ values shown on the horizontal axis, for $Q=1.3$ (top), 4 (lower left) and 10 GeV (lower right subfigure). The $L_2$ sensitivity is approximately equal to the variation in $\chi^2_E$ for experiment $E$ when $A_2^{\mbox{\tiny eff}}[F_2^p(x_B,Q^2)]$ is increased by the 68\% c.l. Hessian uncertainty above its best-fit value at the specified $x_B$. In other words, we increase $A_2^{\mbox{\tiny eff}}[F_2^p(x_B,Q^2)]$ to the upper boundary of the dark error band for the respective $Q$ in Fig.~\ref{fig:OverTheRainbow} and ask how $\chi^2_E$ changes under this variation.

The curves in Fig.~\ref{fig:OTRSensitivity} are for several experiments in the CT18 NNLO fit that show the highest sensitivity to $A_2^{\mbox{\tiny eff}}[F_2^p(x_B,Q^2)]$ at $x_B>0.6$. At $Q=1.3$ GeV, these are BCDMS $ep$ and $ed$ DIS cross sections \cite{Benvenuti:1989fm,Benvenuti:1989rh}, the NMC ratio of $ep$ and $ed$ DIS cross sections \cite{Arneodo:1996qe}, CDHSW $F_2$ and $F_3$ measurements for charged-current DIS on a heavy nucleus \cite{Berge:1989hr}, E866/NuSea $pp$ Drell-Yan cross sections \cite{Webb:2003ps}, and CMS jet production cross sections at 7 and 8 TeV \cite{Chatrchyan:2014gia,Khachatryan:2016mlc}. Although the coverage by these data in the CT18 fit ends roughly at $x\approx 0.75$, they predict $A_2^{\mbox{\tiny eff}}$ at larger $x$ values through extrapolation. 

While there is a reasonable agreement between the experiments on the value of $A_2^{\mbox{\tiny eff}}[F_2^p(x_B,Q^2)]$, the uncertainty bands on $A_2^{\mbox{\tiny eff}}[F_2^p(x_B,Q^2)]$  in Fig.~\ref{fig:OverTheRainbow} emerge as a compromise between the opposite pulls of the contributing experiments. At $Q=1.3$ GeV and $x_B>0.85$ in the left Fig.~\ref{fig:OTRSensitivity}, we observe that the positive variation of $A_2^{\mbox{\tiny eff}}[F_2^p(x_B,Q^2)]$ leads to a decrease of $\chi^2_E$ for the BCDMS $ep$ DIS and E866/NuSeA $pp$ Drell-Yan cross sections by up to 10 units, while at the same time it increases $\chi^2_E$ for the BCDMS $ed$ DIS cross section by a comparable amount. At $x_B=0.6-0.7$, the E866/NuSeA data, together with BCDMS $ed$ DIS and CMS jet production data sets, prefer a larger-than-nominal $A_2^{\mbox{\tiny eff}}[F_2^p(x_B,Q^2)]$, while they are opposed by downward pulls on $A_2^{\mbox{\tiny eff}}[F_2^p(x_B,Q^2)]$ from BCDMS $ep$ DIS, CDHSW charged-current DIS, NMC $ep/ed$ ratio, and other measurements.

At $Q=4$ and 10 GeV, the CDHSW data sets play less prominent role, while the combined HERA I+II DIS data set \cite{Abramowicz:2015mha} imposes some constraints.

\begin{figure}[tb]
\centering
\includegraphics[height=.35\textheight]{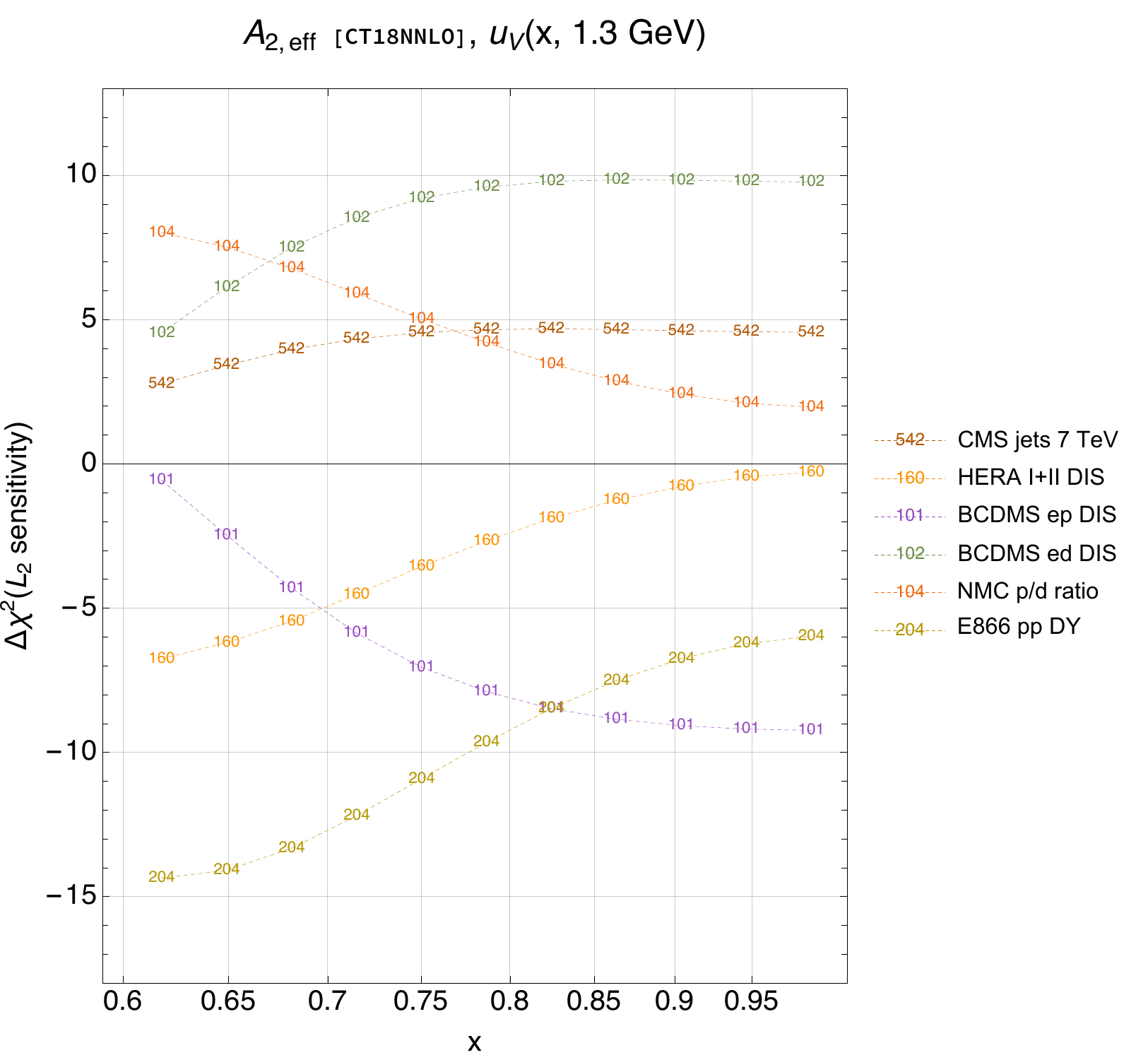}
\includegraphics[height=.35\textheight]{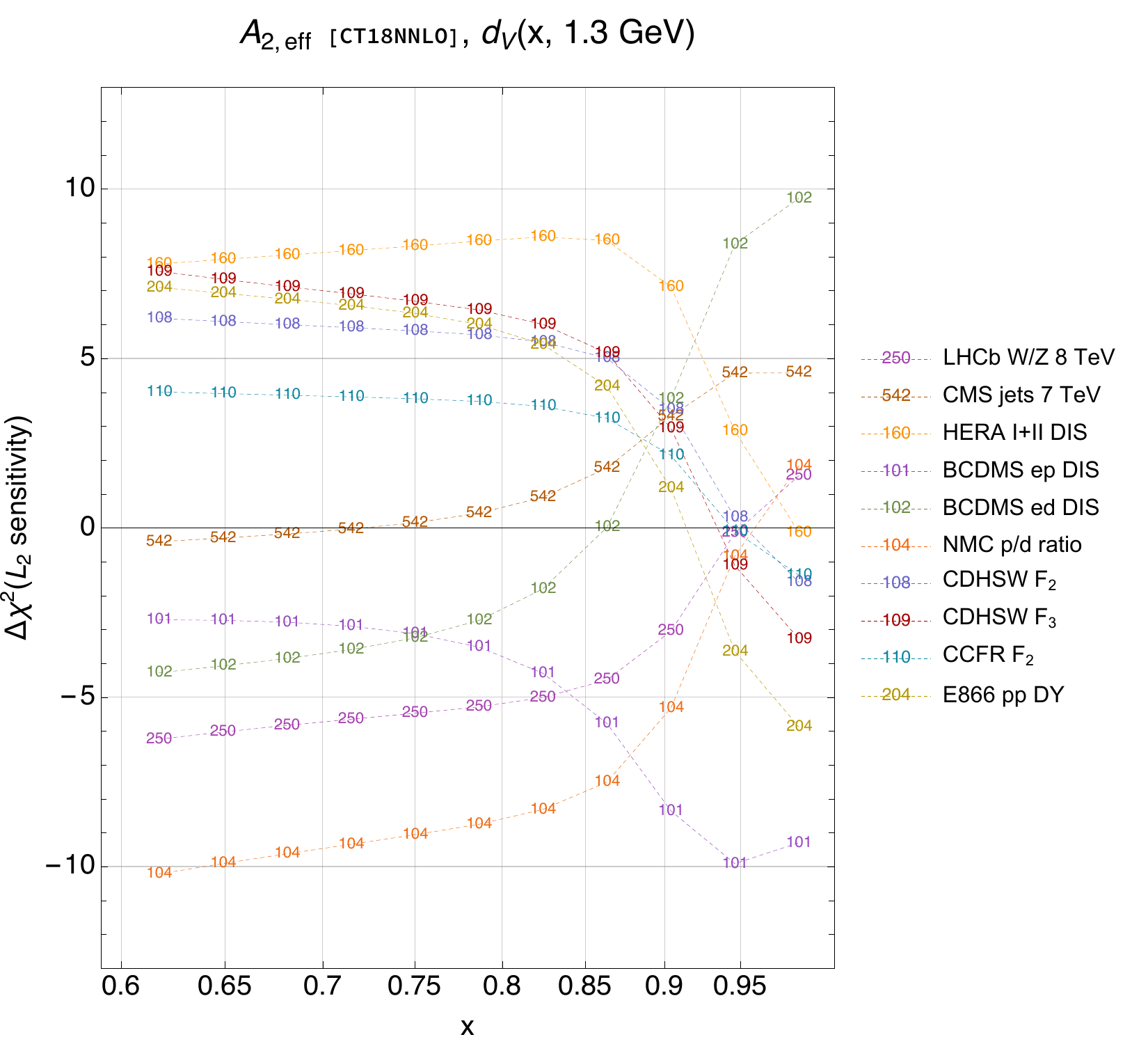}
\caption{ $L_2$ sensitivity to the effective exponent $A_2^{\mbox{\tiny eff}}$ for $u_{\mbox{\tiny V}}(x,Q^2)$ (left) and $d_{\mbox{\tiny V}}(x,Q^2)$ (right) vs. $x$ 
at $Q=1.3$ GeV.}
\label{fig:OTRSensitivityUD}
\end{figure}

Next, we turn to the sensitivities to $A_2^{\mbox{\tiny eff}}$ of $u_{\mbox{\tiny V}}$ and $d_{\mbox{\tiny V}}$ distributions at $Q=1.3$ GeV in Fig.~\ref{fig:OTRSensitivityUD}. The pattern of sensitivities to $A_2^{\mbox{\tiny eff}}$ of $u_{\mbox{\tiny V}}$ in the left panel is visually similar to that for $F_2^p(x_B,Q^2)$ at $Q=4$ GeV in the second Fig.~\ref{fig:OTRSensitivity}. Once again, competing pulls of the BCDMS $ep$ cross sections and E866/NuSea $pp$ Drell-Yan cross sections against the BCDMS $ed$ DIS cross sections stand out at $x>0.8$. At $x=0.6$, the E866/NuSea data and to some extent the HERA DIS data prefer higher $A_2^{\mbox{\tiny eff}}$ than the BCDMS $ed$ and NMC $p/d$ DIS measurements. 

The pattern on the pulls on $A_2^{\mbox{\tiny eff}}$ is more elaborate for $d_{\mbox{\tiny V}}$ in the right panel of Fig.~\ref{fig:OTRSensitivityUD}. The $d_{\mbox{\tiny V}}$ is less constrained at large $x$ than $u_{\mbox{\tiny V}}$ because the relevant data in the global fit 
are dominated by neutral-current DIS measurements that are four times more sensitive to up-type quark PDFs than to down-type ones. In Figs.~\ref{fig:CT18nn_A2u_A2d_ori} and \ref{fig:CT18nn_A2u_A2d}, we observed a moderate PDF uncertainty on $A_2^{\mbox{\tiny eff}}$ for $u_{\mbox{\tiny V}}$ and a much larger uncertainty on $d_{\mbox{\tiny V}}$, especially at $x>0.75$ where the parametrization uncertainty dominates.

Since the published CT18 parametrization sets $A_{2,d_{\mbox{\tiny V}}}=A_{2,u_{\mbox{\tiny V}}}$, in the region $x>0.9$ with no data, $A_2^{\mbox{\tiny eff}}$ for $d_{\mbox{\tiny V}}$ essentially follows that for $u_{\mbox{\tiny V}}$. Namely, its value reflects a tradeoff between the opposing pulls of the BCDMS $ep$ and $ed$ DIS data sets. At $x<0.9$, we see a different pattern, whereby a fairly strong upward pull on $A_2^{\mbox{\tiny eff}}$ by the NMC $p/d$ ratio, complemented by  LHCb $W/Z$ production at 8 TeV \cite{Aaij:2015zlq} and both BCDMS $ep$ and $ed$ DIS, is opposed by the combined HERA DIS, E866/NuSea $pp$, as well as by CDHSW and CCFR \cite{Yang:2000ju} inclusive charged-current DIS data on heavy nuclei. 

In these comparisons, we see some differences between the preferences of scattering experiments on the proton versus deuteron and heavy-nuclei scattering. While these differences do not rise to clear disagreements, they nevertheless suggest importance of the treatment of nuclear effects in future PDF fits. 

\section{Implications for low-energy dynamics}
\label{sec:disc}

We can now address the question of whether phenomenological PDFs reflect manifestations of low-energy dynamics, based on the discussion of the physical meaning of PDFs in nonperturbative approaches and phenomenological fits in Sec.~\ref{sec:LargeXRealQCD}, the mathematical arguments of Sec.~\ref{sec:Mimicry}, and numerical results in Secs.~\ref{sec:a2effF2p}-\ref{sec:a2effByProcess}.
We will focus on two  points relevant for further analyses: the relation between the nonperturbative approaches and phenomenological PDFs, and studies of PDFs in the pion.

    \subsection{On the relation to nonperturbative approaches}
It seems appropriate to assume that the quark counting rules are realized when the QCD coupling in semi-hard scattering is reasonably small, so that the final Fock states are dominated by contributions with a small number of perturbative partons interacting through nearly perturbative interactions. The reality is more complex: radiation in the PQCD regime results in logarithmic evolution of the partonic probability from large $x$ at low $Q^2$ towards smaller $x$ values at higher $Q^2$, which in turn introduces $Q^2$ dependence of the $(1-x)$ exponents quantified by their anomalous dimensions \cite{Goldberger:1976vp,Soper:1976jc,Ball:2016spl}. The ideal window in $\{x, Q^2\}$ for QCR studies overlaps with the resonance region in DIS, from which the structure functions can still be extracted, but the global fits of PDFs in this domain must include target-mass corrections --- a kinematic effect dependent on a specific scattering process --- and other higher-twist terms~\cite{Accardi:2016qay}. Threshold resummation \cite{Aicher:2010cb, Bonvini:2015ira} and related nonperturbative effects, such as the modified running of the QCD coupling constant \cite{Courtoy:2013qca}, are important at the highest $x$.

Reconciliation of the nonperturbative and phenomenological definitions of PDFs runs into important differences between the degrees of freedom adopted in various theoretical approaches. In the PQCD collinear factorization framework exemplified by Eq.~(\ref{eq:DISfactorization}) for deep inelastic scattering, universal PDFs correspond to the long-distance part of the hadronic cross section that is perturbatively expanded as a series of the small QCD coupling and power-suppressed (twist) terms. 
Here, both expansions are made possible by the presence of a scale $Q > 1$ GeV in the hard cross section. Phenomenological PDFs are defined in an $\overline{MS}$ or another factorization scheme introduced to separate long- and short-distance radiative contributions. 

Nonperturbative predictions for the hadron structure do not have an inherent large energy scale that sets the small expansion parameters. They describe the internal structure at a low hadronic scale $\mu_0 < 1$ GeV and must be matched to the factorized PQCD predictions at an intermediate scale  $Q_0>\mu_0$. The bridge between these two scales, namely the one-to-one connection between the low-scale dynamic degrees of freedom to the PQCD quarks and gluons, remains an unsolved problem, with hints available in Dyson-Schwinger approaches and lattice QCD, {\it e.g.} in ~\cite{Aguilar:2008xm,Roberts:1994dr,Binosi:2014aea,Bogolubsky:2009dc}. 

In the absence of such a clear connection at present, one might resort to a model of the spin and flavor dependence of the whole operator matrix element in the $\overline{MS}$ definition (\ref{eq:MSbarPDF}) of $f_{a/p}(x,Q^2)$. Results obtained with an $SU(6)$-symmetric wave function or an $SU(6)$-broken~\cite{Close:1973xw}, quark-diquark configuration~\cite{Carlitz:1975bg,Carlitz:1976in,Kaur:1977ce,Jakob:1997wg}, to name a few, reveal useful hints about the 
 $x$ dependence of PDFs at large $x$. 
 We relied on these considerations when equating $A_2$, the $(1-x)$ exponents of the valence up and down quarks, in the PDF parametrizations adopted in the CT18 analysis. 
 
A similar spin-flavor consideration for the pion, discussed in the next subsection, seems less relevant at mild energies due to the pion's pseudo-Nambu--Goldstone origin. 

    \subsection{ On the pion case \label{sec:pion}}
    
In Refs.~\cite{Ezawa:1974wm,Soper:1976jc,Farrar:1975yb,Berger:1979du}, the counting rules were also formulated for the structure function of the pion, predicting a $(1-x)^2$ falloff near the threshold.  This behavior is often predicted based on the expression of the pion PDF, or distribution amplitude, in terms of the long-distance pion wave function  $\phi$ and semi-hard scattering contribution dominated by the $q\bar{q}$ state, as reviewed in Sec.~\ref{subsec:ff} and Fig.~\ref{fig:QCRdiagrams}.

The pion structure provides a fascinating window on QCD dynamics. Kinematics of the target meson in neutral-current DIS (Sec.~\ref{sec:LargeXRealQCD}) takes a new meaning in light of the pion mass generation, a key emerging feature for pion-related observables. Chiral symmetry and its breaking govern the pion structure at low- to mid-energies. The nonperturbative quark-quark interaction cannot be replaced by a hard-gluon exchange at energies at which manifestations of chiral symmetry are substantial compared to PQCD interactions. This point is also  highlighted for a related case of  hard exclusive processes in Ref.~\cite{Shuryak:2020ktq}. The latter are best understood by comparing a fully nonperturbative approach for predicting the pion electromagnetic form factor to a large-$Q^2$ description in terms of distribution amplitudes and a hard-scattering part~\cite{Lepage:1979zb,Lepage:1980fj}.  The addition of nonperturbative effects to the hard-gluon exchange in \cite{Shuryak:2020ktq} improves the description of the form factor at low/moderate $Q^2$ and provides a better transition to the  asymptotic "perturbative" behavior \cite{Lepage:1979zb} associated with the QCRs for $Q^2$ up to a least 10 GeV$^2$. The role played by the large-$x$ distribution amplitude (or the structure function) in the behavior of the pion electromagnetic form factor at large $Q^2$ has been emphasized numerous times, see, {\it e.g.}~\cite{Melnitchouk:2002gh,Chang:2013nia,Chen:2018rwz}. In other words, for the pion, the concepts of  weak coupling and a loosely-bound initial state assumed in the QCR picture cannot be approached without also considering the long-distance effects induced by chiral symmetry.
\\

Present and future experiments -- at JLab, EIC, or AMBER/COMPASS++ -- aim to unveil the pion structure in DIS and Drell-Yan pair production. The questions examined throughout this manuscript apply to the fits of pion PDFs. Given the simpler valence structure of the pion and the pion's low mass, we anticipate considerable simplifications with respect to the case of the proton. The main experimental constraints on the pion PDFs at large $x$ for now come from the E615 Drell-Yan pair production in pion-nucleus scattering~\cite{Conway:1989fs}. In this process, a large momentum fraction $x_1$ for the pion corresponds to a small $x_2$ for a nucleus, except in the true threshold limit when $s\approx Q^2$, where no measurements currently exist. In such kinematic regime, when the nuclear beam remnant creates high hadronic multiplicities in the final state, one must carefully revisit the justifications for PQCD factorization for the Drell-Yan process presented at the end of Sec.~\ref{sec:LargeXRealQCD}. Nuclear shadowing in the initial state and interactions with the nuclear remnant in the final state may elevate the power-suppressed contributions as compared to the nucleon scattering. A concern about having a genuinely free pion target arises in  prompt photon production in pion-proton scattering, as well as in leading neutron electroproduction. On the positive side, the finely binned E615 data points extend to $x_1=0.99$, very close to the end point. Due to the smallness of the pion mass, target-mass corrections for pion DIS are almost negligible~\cite{Melnitchouk:2002gh}.

Modern global analyses for the pion PDFs~\cite{Barry:2018ort,Novikov:2020snp,Bourrely:2020izp} now apply advanced theoretical frameworks, {\it e.g.} threshold resummation~\cite{Aicher:2010cb,Barry}. 
Depending on the theoretical framework, the recent analyses find that the pion data are compatible with a nominal $(1-x)$ or $(1-x)^2$ behavior of the PDF parametrization at $Q_0$.
When the large-$x$ resummation is included for the DY data, the authors \cite{Aicher:2010cb,Barry} find a fast falloff of the valence pion PDFs in $(1-x)$, consistent with $A_2=2$. 
While threshold resummation must come into play at large $x$, it is not a sufficient condition for testing that the extracted PDFs fulfill the QCR predictions. None of these analyses addresses the functional mimicry of high-degree polynomial fits discussed in  Sec.~\ref{sec:Mimicry}.

As we emphasized in that section, without knowing the exact functional form of the PDFs, one must include the end-point region, ideally $x>0.9$, to pin down the low powers of $(1-x)$ in the monomial expansion. Physical uncertainties grow in the end-point region. For example, if threshold resummation is necessary, one must account for its uncertainties due to the choice of factorization scales, matching on the fixed-order prediction, and power-suppressed terms.

On the other hand, a fit that is restricted to smaller values of $x$ introduces a spurious correlation between the coefficients with low and high powers of $(1-x)$. This correlation depends on the fitted data sample and strongly modifies the lowest-power monomial terms, as illustrated in Figs.~\ref{fig:Beziers} (b-d).

An alternative approach in Sec.~\ref{sec:A2eff} computes the effective exponent $A_2^{\mbox{\tiny eff}}$ that can be compared against theoretical predictions without reconstructing the analytic form of the PDFs. The value of $A_2^{\mbox{\tiny eff}}$ depends on the range of $x$. Its {\it global} trend, reflecting slow variations over $x$, must be distinguished from the {\it local} one, existing in a small neighborhood of the examined data bins. For complex PDF parametrizations, like the ones used by NNPDF, the effective exponent may have large local variations, due to mimicry, and require averaging over PDF replicas and/or a range of $x$ in order to determine its $x\to 1$ limit. 
For smooth PDF parametrizations, like the CT18 or JAM ones, averaging is not necessary: the $A_2^{\mbox{\tiny eff}}$ values computed based on such PDFs follow smooth trends and can be determined for comparisons against the QCR's at the momentum fractions of about $0.8$, as has been done in Fig.~\ref{fig:OverTheRainbow}. 

 Analogous considerations apply to PDFs predicted by low-energy models. Ideally, these models should provide uncertainty bands for the cross sections or PDFs that can be verified (or falsified) by the experimental data {\it over the full $x$ range}. Such uncertainties are difficult to estimate faithfully. In the absence of the uncertainty bands, one must focus on the aspects of the low-energy predictions that are preserved in hard scattering. For example, a low-energy dynamic effect like the broadening of the parton distributions due to the emergence of dynamical mass \cite{Davidson:1994uv,Petrov:1998kg,Bednar:2018mtf,Ding:2019lwe} may favor a particular $(1-x)$ falloff power, yet consistency of this power with the experimental data in some kinematic region is not sufficient for validating such prediction as the only viable one. The local trend quantified by the empirical $A_2^{\mbox{\tiny eff}}$ values may differ from the global trend in complex nonperturbative models. A  comparison against the QCRs requires experimental access to the end point $x=1$, where additional dynamical effects are most pronounced. 

Secs.~\ref{sec:a2effF2p} and \ref{sec:a2effPDF} demonstrate that $A_2^{\mbox{\tiny eff}}$ depends on the factorization scale $Q$. The pion valence PDFs obey the same non-singlet evolution equation as the proton ones, thus $A_2^{\mbox{\tiny eff}}$ for the pion PDFs may change by 0.5-1 units within the typical $Q$ range, like in the proton case. The evolution of a PDF that fulfils the QCRs at a low, pre-factorization scale may either increase or decrease $A_2^{\mbox{\tiny eff}}$ at a higher $Q$, reflecting the functional mimicry.

\section{Conclusions}

Complementary to the constraints based on first principles, the quark counting rules offer predictions for DIS structure functions and PDFs near the elastic threshold. In this picture, when a hadron target is almost unperturbed, and the strong coupling constant is small, the structure functions in a nucleon exhibit a $(1-x)^p$ fall-off in the limit $x\to 1$, reflecting exchanges of semihard gluons between the quarks in the incoming bound state, as discussed in Sec~\ref{sec:QCRweakconstant}. 

We examined two possible strategies for testing the quark counting rules with experimental data. From a purely mathematical point of view, the concept of polynomial mimicry, demonstrated in Sec.~\ref{sec:Mimicry} by employing the B\'ezier curve technique, reveals a limitation in reconstructing the exact functional forms of PDFs from discrete data, whether based on an interpolation or a fit. It is not possible to uniquely determine the powers of a $(1-x)$ monomial expansion except very closely to the end point. Associated uncertainties can be very large.

As an alternative, Sec.~\ref{sec:A2eff} showed that it is possible to define an {\it effective} exponent to examine the large-$x$ behavior of any functional form for the PDFs.
The leading-power structure function $F_2^p(x, Q^2)$ reconstructed within the CT18 NNLO global analysis \cite{Hou:2019efy} agrees, within error bands at moderate scales $Q$,  with the predicted power law. However, a non-negligible shift in the effective $(1-x)$-exponent with increasing $Q^2$ is observed in Fig.~\ref{fig:OverTheRainbow}, as expected from DGLAP evolution. Moreover, the resonance region in DIS forbids a reliable analysis at large-$x$ values and photon virtuality of a few GeVs.

This has led us to our  bottom line: exploration of the relevance of quark counting rules for high-energy processes must address various factors arising from both theory and statistics. We have investigated the pertinent issues in the case of PDFs for the nucleon, for which the situation is best understood. These include differences between the quark counting rules for hadronic observables, such as $F_2(x,Q^2)$, as opposed to $\overline{MS}$ PDFs; the universality of the PDFs and the role of power-suppressed and soft corrections; the phenomenological PDF uncertainty reflecting the scarcity of the data at large $x$ as well as the choice of the PDF parametrization. The latter point has been developed by studying $N=363$ replicas of the CT18NNLO analysis, see Sec.~\ref{sec:CT18}. 

Global analyses are based on experimental data from several processes, such as DIS and DY. When considering the effective power laws at the PDF level in Sec.~\ref{sec:a2effPDF}, we have found that not all flavors behave on the same footing. The valence up distribution is consistent with the running exponent of three, the valence down and gluon distributions on average tend to have lower-than-expected exponents, the sea quark exponent is too high, see Fig.~\ref{fig:CT18nn_A2u_A2d_ori}. The ${\bar u}+{\bar d}$ exponent only slightly decreases with $Q^2$. In the same spirit, the preferred effective exponents depend at some level on the fitted experimental process, as highlighted in Sec.~\ref{sec:a2effByProcess}. 

In conclusion, we emphasize that the quark counting rules emerge in the limit of weak coupling in processes with little underlying hadronic activity. Violations of these conditions in at least some high-energy processes put in question the universality of the rules, which is especially relevant for uses in global analyses. On the other hand, there may be experimental measurements that favor the QCRs, such as a subclass of DIS events with low final-state hadronic multiplicities. The other possibility is offered by pion scattering discussed in Sec.~\ref{sec:pion}, which is less affected by collateral factors present in the nucleon or nuclear cases.

An experimental observation of the $x$ dependence predicted by a nonperturbative calculation constitutes an insufficient, but non-redundant condition for validating the calculation. Functional forms of fitted PDFs are unnecessary but sufficient for describing the data. Only by measuring the structure functions/PDFs near the end point $x=1$ one may reveal evidence of the primordial power law, as we have demonstrated in Sec.~\ref{sec:Mimicry} based on the comparison   of the monomial and B\'ezier expansions.  Reconciliation of the predictive phenomenological fits and the interpretative nonperturbative approaches within uncertainties will require more data at large $x$  as well as efforts to address theoretical issues that hinder our understanding of cross sections at the end point $x=1$.

\section*{Acknowledgements}

We thank  A. Accardi, N. Boileau-Despr\'eaux, R. Ent, T.-J. Hobbs, S. Liuti, W.~Melnitchouk, F. I. Olness, M.~V.~Polyakov, C.~D.~Roberts, and CTEQ-TEA collaborators for stimulating discussions and inspirations.
AC is supported by UNAM Grant No. DGAPA-PAPIIT IA101720 and CONACyT Ciencia de Frontera 2019 No.~51244 (FORDECYT-PRONACES). PN is partially supported by the U.S.~Department of Energy under Grant No.~DE-SC0010129.

\bibliographystyle{apsrev4-1}
\bibliography{ct18bibtex}

\end{document}